\documentclass[aps,prd,showpacs,eqsecnum,twocolumn,superscriptaddress,nofootinbib]
{revtex4-1}
\usepackage{amsmath,amssymb,graphicx,color}

\begin{document}

\title{Frequency-domain gravitational waveform models for inspiraling binary neutron stars}

\author{Kyohei Kawaguchi}\affiliation{Max Planck Institute for Gravitational Physics (Albert Einstein Institute), Am M\"{u}hlenberg 1, Potsdam-Golm, 14476, Germany}\affiliation{Center for Gravitational Physics,
  Yukawa Institute for Theoretical Physics, 
Kyoto University, Kyoto, 606-8502, Japan} 

\author{Kenta Kiuchi}\affiliation{Center for Gravitational Physics,
  Yukawa Institute for Theoretical Physics, 
Kyoto University, Kyoto, 606-8502, Japan}

\author{Koutarou Kyutoku} \affiliation{
Theory Center, Institute of Particle and Nuclear Studies, KEK,
Tsukuba 305-0801, Japan\\
Department of Particle and Nuclear Physics, the Graduate University
for Advanced Studies (Sokendai), Tsukuba 305-0801, Japan\\
Interdisciplinary Theoretical Science (iTHES) Research Group, RIKEN,
Wako, Saitama 351-0198, Japan
}\affiliation{Center for Gravitational Physics,
  Yukawa Institute for Theoretical Physics, 
Kyoto University, Kyoto, 606-8502, Japan} 

\author{Yuichiro Sekiguchi} \affiliation{Department of Physics, Toho
  University, Funabashi, Chiba 274-8510, Japan}

\author{Masaru Shibata}\affiliation{Center for Gravitational Physics,
  Yukawa Institute for Theoretical Physics, 
Kyoto University, Kyoto, 606-8502, Japan}
\affiliation{Max Planck Institute for Gravitational Physics (Albert Einstein Institute), Am M\"{u}hlenberg 1, Potsdam-Golm, 14476, Germany}

\author{Keisuke Taniguchi}
\affiliation{Department of Physics, University of the Ryukyus, Nishihara, Okinawa 903-0213, Japan}
\date{\today}

\newcommand{\beq}{\begin{equation}}
\newcommand{\eeq}{\end{equation}}
\newcommand{\beqn}{\begin{eqnarray}}
\newcommand{\eeqn}{\end{eqnarray}}
\newcommand{\pa}{\partial}
\newcommand{\vp}{\varphi}
\newcommand{\varep}{\varepsilon}
\newcommand{\ep}{\epsilon}
\newcommand{\comp}{(M/R)_\infty}
\newcommand{\rednote}[1]{{\color{red} (#1)}}
\begin{abstract}

We develop a model for frequency-domain gravitational waveforms from inspiraling binary neutron stars. Our waveform model is calibrated by comparison with hybrid waveforms constructed from our latest high-precision numerical-relativity waveforms and the SEOBNRv2T waveforms in the frequency range of $10$--$1000\,{\rm Hz}$. We show that the phase difference between our waveform model and the hybrid waveforms is always smaller than $0.1\, {\rm rad}$ for the binary tidal deformability, ${\tilde \Lambda}$, in the range $300\lesssim{\tilde \Lambda}\lesssim1900$ and for the mass ratio between 0.73 and 1. We show that, for $10$--$1000\,{\rm Hz}$, the distinguishability for the signal-to-noise ratio $\lesssim50$ and the mismatch between our waveform model and the hybrid waveforms are always smaller than 0.25 and $1.1\times10^{-5}$, respectively. The systematic error of our waveform model in the measurement of ${\tilde \Lambda}$ is always smaller than $20$ with respect to the hybrid waveforms for $300\lesssim{\tilde \Lambda}\lesssim1900$. The statistical error in the measurement of binary parameters is computed employing our waveform model, and we obtain results consistent with the previous studies. We show that the systematic error of our waveform model is always smaller than $20\%$ (typically smaller than $10\%$) of the statistical error for events with the signal-to-noise ratio of $50$.
\end{abstract}


\maketitle

\section{Introduction}
On 17th of August 2017, three ground-based gravitational-wave detectors, advanced LIGO~\cite{TheLIGOScientific:2014jea} and advanced Virgo~\cite{TheVirgo:2014hva}, reported the first detection of gravitational waves from a binary neutron star merger referred to as GW170817~\cite{Abbott2017}. One of the monumental achievements for this detection is the measurement of the tidal deformability of neutron stars. Gravitational waves from binary neutron stars contain rich information of the neutron stars, in particular, the information of their masses and quantities related to equation of state. The simultaneous measurement of these quantities of the neutron stars provides a substantial constraint on the equation of state of nuclear matter which is yet poorly understood~\cite{Lattimer:2012nd}. Among various proposals, the tidal deformability of neutron stars has been proposed as one of the most promising quantities related to the equation of state that can be extracted from the gravitational-wave observation~\cite{Lai:1993pa,Mora:2003wt,Flanagan:2007ix,Read:2009yp,Damour:2009wj,Hinderer:2009ca,Vines:2011ud,Damour:2012yf,Bini:2012gu,Favata:2013rwa,Yagi:2013baa,Read:2013zra,Bini:2014zxa,Bernuzzi:2014owa,Wade:2014vqa,Lackey:2014fwa}. By the observation of GW170817, it is confirmed that the measurement of the neutron-star tidal deformability is indeed possible. While various equations of state are still consistent with the measurement of the tidal deformability for this event, a number of detections of gravitational waves from binary neutron stars by the advanced detectors~\cite{TheLIGOScientific:2014jea,TheVirgo:2014hva,Kuroda:2010zzb} are expected in the next few years~\cite{Kalogera:2006uj,Abadie:2010cf,Kim:2013tca,Abbott2017}, and the measurement of neutron-star properties from them will surely give a great impact on both astrophysics and nuclear physics~\cite{Agathos:2015uaa}.

To extract the tidal deformability of neutron stars from the observed gravitational-wave data, an accurate theoretical waveform template is crucial. For deriving the waveform models, many efforts have been made. For the early inspiral stage, the waveforms including the linear-order tidal effects are derived by post-Newtonian (PN) calculation. The Newtonian terms are first derived by~\cite{Flanagan:2007ix}, and the 1PN terms by Vines, Flanagan and Hinderer~\cite{Vines:2011ud}. However, it is shown in Refs.~\cite{Favata:2013rwa,Yagi:2013baa,Lackey:2014fwa,Wade:2014vqa} that theses waveforms are not accurate enough for the estimation of the tidal deformability, because of the presence of a significant systematic error due to the unknown higher-order PN terms. In particular, the lack of higher-order PN terms in the point-particle part of gravitational waves is problematic since the tidal effects are only significant in the last part of the inspiral stage for $f\gtrsim400\,{\rm Hz}$~\cite{Hinderer:2009ca,Damour:2012yf}, where $f$ is the gravitational-wave frequency. To incorporate higher-order PN effects, Damour and his collaborators derived the waveforms employing the {\it effective-one-body} (EOB) formalism including the tidal effects up to the 2.5 PN order~\cite{Damour:2009wj,Bini:2012gu,Damour:2012yf,Bini:2014zxa,Bernuzzi:2014owa}. In the EOB formalism, higher-order PN correction is included by re-summation techniques and calibrated by comparing the model waveforms with those derived by numerical-relativity simulations of binary black holes. Hinderer and her collaborators have pushed these works further and derived the EOB waveforms considering dynamical tides~\cite{Hinderer:2016eia,Steinhoff:2016rfi,Dietrich:2017feu}. It is shown that these latest tidal-EOB (TEOB) waveforms can be accurate even up to $\approx3\,{\rm ms}$ before the onset of merger~\cite{Kiuchi:2017pte}. However, the phase difference between the TEOB waveforms and the numerical-relativity results is still larger than $\approx 1\,{\rm rad}$ after two neutron stars come into contact for the case that the neutron-star radii are larger than $\approx13\,{\rm km}$. Thus, further improvement of the waveform model is needed to suppress the systematic error in the measurement of the tidal deformability.

High-precision numerical-relativity simulation is the unique method to predict the tidal effects in a regime where the non-linear effect of hydrodynamics should be taken into account in the framework of general relativity~\cite{Thierfelder:2011yi,Baiotti:2011am,Bernuzzi:2012ci,Radice:2013hxh,Hotokezaka:2015xka,Haas:2016cop,Hotokezaka:2016bzh,Dietrich:2017feu,Dietrich:2017aum,Kiuchi:2017pte}. Recently, because of the progress of simulation technique and increase of the available computational resources, the precision and duration of the numerical-relativity waveforms have been remarkably improved. In particular, the waveforms for more than $15$ inspiral orbits are derived with a sub-radian order error in our previous study~\cite{Kiuchi:2017pte}. Although our work provides one of the longest numerical-relativity waveforms for inspiraling binary neutron stars to date, they are still too short for the use of constructing an accurate waveform model. Hybrid waveforms employing analytic waveforms for the low-frequency part and numerical-relativity waveforms for the high-frequency part are used to solve this problem~\cite{Read:2013zra,Hotokezaka:2016bzh}.

In this paper, we develop an accurate model for gravitational waves from inspiraling binary neutron stars taking tidal deformation of neutron stars into account. We calibrate our waveform model employing hybrid waveforms constructed from our latest numerical-relativity waveforms and the TEOB waveforms. The waveform model is derived in the frequency domain as in the Phenom-series for binary black holes~\cite{Khan:2015jqa} for convenience in data analysis. We note that a gravitational waveform model for binary neutron stars based on numerical-relativity waveforms is also derived in Ref.~\cite{Dietrich:2017aum} in a similar manner. The main difference between our and their works is the difference of the numerical-relativity waveforms and the TEOB waveforms used for the model calibration. Moreover, in Ref.~\cite{Dietrich:2017aum}, the waveform model is derived in the time domain, and then, is transformed to a frequency-domain waveform model employing the stationary-phase approximation, while our waveform model is calibrated directly in the frequency domain. We present a comparison between the model of Ref.~\cite{Dietrich:2017aum} and our model in Appendix~\ref{appE}.

This paper is organized as follows: In Sec.~\ref{sec2}, we summarize the waveforms used for deriving and calibrating our waveform model, and present the method to derive our waveform model. In Sec.~\ref{sec3}, we examine the validity of our waveform model derived in Sec.~\ref{sec2} by computing the distinguishability and the systematic error in the measurement of binary parameters using the hybrid waveforms as hypothetical signals. In Sec.~\ref{sec4}, we compute the statistical error in the measurement of the binary parameters based on the standard Fisher-matrix analysis. We present the summary of this paper in Sec.~\ref{sec5}. Unless otherwise stated, we employ the units of $c=G=1$, where $c$ and $G$ are the speed of light and the gravitational constant, respectively.

\section{Model}\label{sec2}
In this section, we derive a frequency-domain waveform model for gravitational waves from inspiraling binary neutron stars. The Fourier spectrum of gravitational waves from a binary neutron star, ${\tilde h}\left(f\right)$, can be written in terms of the amplitude, $A(f)$, and phase, $\Psi(f)$, as
\begin{align}
	{\tilde h}\left(f\right)=A\left(f\right)e^{-i\Psi\left(f\right)}.
\end{align}
For binary neutron stars, both phase and amplitude of the gravitational-wave spectrum depend on tidal deformation of neutron stars. We define the tidal part of the gravitational-wave phase by\footnote{See Sec.~\ref{sec2b} for the ambiguity in this definition due to the time and phase shifts.}
 \begin{align}
	\Psi_{\rm tidal}\left(f\right)&=\Psi\left(f\right)-\Psi_{\rm pp}\left(f\right),
\end{align}
where $\Psi_{\rm pp}\left(f\right)$ is the gravitational-wave phase of a binary black hole with the same mass as the binary neutron star (hereafter referred to as the point-particle part of the phase). Similarly, the tidal part of the gravitational-wave amplitude is defined by
\begin{align}
	A_{\rm tidal}\left(f\right)=A\left(f\right)-A_{\rm pp}\left(f\right),
\end{align}
 where $A_{\rm pp}\left(f\right)$ is the gravitational-wave amplitude of a binary black hole with the same mass as the binary neutron star (hereafter referred to as the point-particle part of the amplitude). In this work, we employ the SEOBNRv2 waveforms~\cite{Taracchini:2013rva} as the fiducial point-particle part of gravitational waves. This is because we employ the  SEOBNRv2T waveforms for the low-frequency part of the hybrid waveforms (see Sec.~\ref{sec2a}), and the point-particle limit of the  SEOBNRv2T formalism agrees with the SEOBNRv2 formalism.

In the following subsections, the tidal-part models for the gravitational-wave phase and amplitude are derived. First, we derive a frequency-domain model for the hybrid waveforms focusing only on equal-mass binary cases. Then, we extend our study to unequal-mass binary cases.

We also derive simple analytic point-particle part models for both phase and amplitude of gravitational waves that reproduce the SEOBNRv2 waveforms with reasonable accuracy for the total mass in the range of $2.4$--$3.0\,M_\odot$ and for the symmetric mass ratio in the range of $0.244$--$0.25$. We employ these point-particle models for the analysis in Sec.~\ref{sec3} and Sec.~\ref{sec4}. The details and the derivation of these point-particle models are presented in Appendix~\ref{appA}. 

\begin{figure}
 	 \includegraphics[width=.95\linewidth]{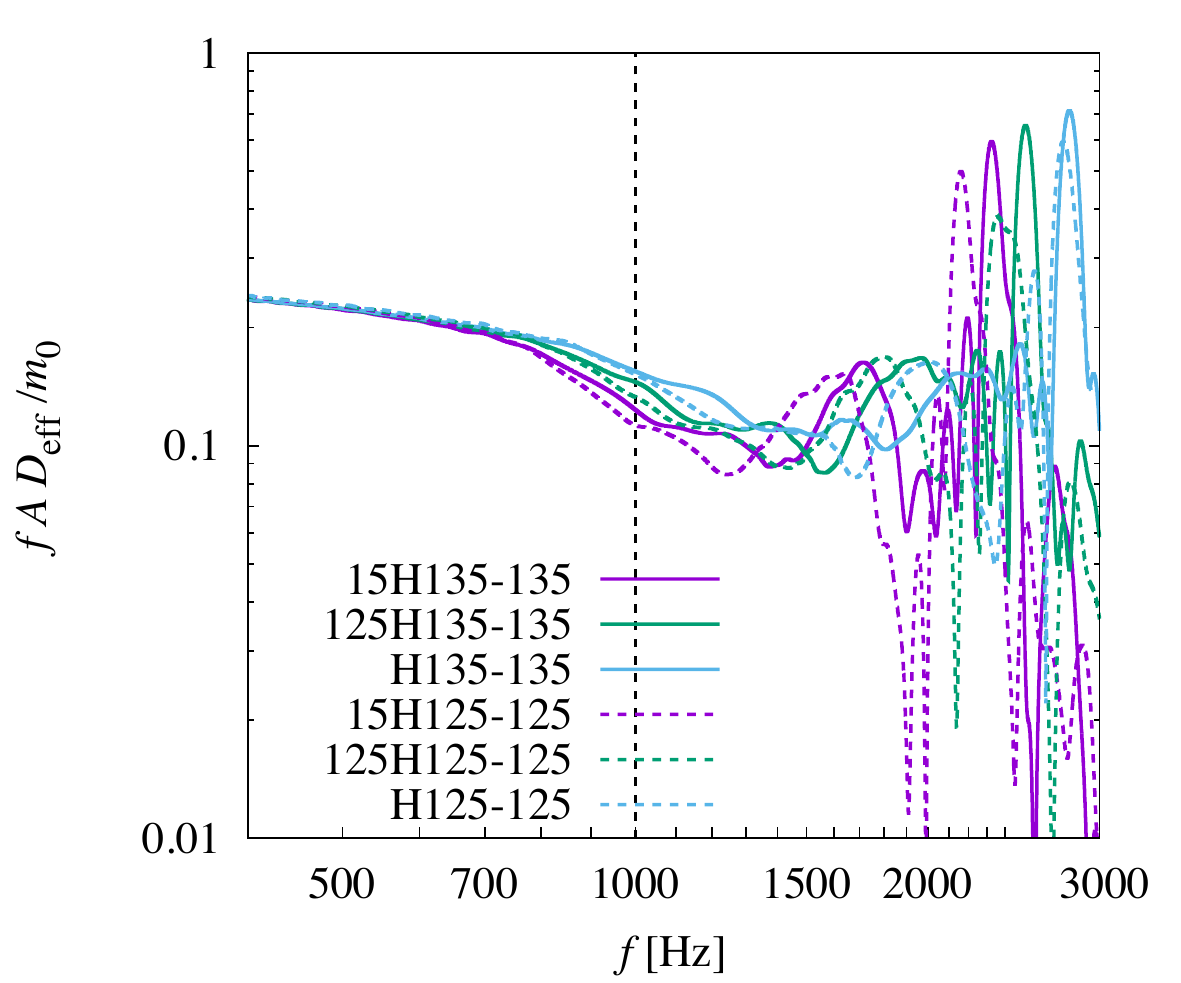}
 	 \caption{The comparison of Fourier spectra of gravitational waves from binary neutron stars with three different equations of state and with two values of total mass. $A(f)$, $D_{\rm eff}$, and $m_0$ denote the amplitude of the spectrum, the effective distance to the gravitational-wave source, and the total mass of the binary, respectively. The vertical dashed line denotes $f=\,1000\,{\rm Hz}$.}\label{fig:amp_135_135}
\end{figure}

We note that, in this work, we focus only on gravitational waves for $f\le1000\,{\rm Hz}$. The reason for this is that the gravitational-wave spectra for $f>1000\,{\rm Hz}$ would be affected by the post-merger waveforms: In Fig.~\ref{fig:amp_135_135}, we show the amplitude of the gravitational-wave spectra for several binary neutron star models (see Sec.~\ref{sec2a} for the details of binary neutron star models). Figure~\ref{fig:amp_135_135} shows that the amplitude is no longer a monotonic function of the gravitational-wave frequency for $f\gtrsim1100\,{\rm Hz}$. This suggests that both amplitude and phase of the spectra are affected by the waveforms after the merger that can be modified by detailed physical effects (see Appendix~\ref{appB} for a detailed analysis). Thus, we have to restrict our attention to the frequency of $f \le1000\,{\rm Hz}$. In this work, we also focus only on the case that the spins of neutron stars are absent. We leave the extension of our waveform model for the future task.

\subsection{Time-domain hybrid waveforms}\label{sec2a}

\begin{table*}[t]
\centering
\caption{Model name, mass of each neutron star, $m_i$ $(i=1,2)$, equations of state (EOS) employed, chirp mass, ${\cal M}_{\rm c}=\left(m_1m_2\right)^{3/5}/\left(m_1+m_2\right)^{1/5}$, symmetric mass ratio, $\eta=m_1m_2/\left(m_1+m_2\right)^2$, binary tidal deformability ${\tilde \Lambda}$ (see Eq.~\eqref{eq:symlambda} for its definition), location of outer boundaries along each axis denoted by $L$, and finest grid spacing of the simulation, $\Delta x_{\rm finest}$. The radius and the dimensionless tidal deformability of neutron stars for each equation of state are listed in Table~\ref{tb:eos_model}.}
\begin{tabular}{c|ccc|ccc|cc}
\hline\hline
Model & $m_1$ & $m_2$ &EOS	& ${\cal M}_{\rm c}$&$\eta$& ${\tilde \Lambda}$&~~~$L\,[{\rm km}]$~~~ & $\Delta x_{\rm finest}\,[{\rm m}]$\\\hline
15H135-135	&	1.35	&	1.35	&	15H		&1.17524&0.25&	1211	&7990&86\\
125H135-135	&	1.35	&	1.35	&	125H	&1.17524&0.25&	863		&7324&79\\
H135-135	&	1.35	&	1.35	&	H		&1.17524&0.25&	607		&6991&75\\
HB135-135	&	1.35	&	1.35	&	HB		&1.17524&0.25&	422		&6392&69\\
B135-135	&	1.35	&	1.35	&	B		&1.17524&0.25&	289		&5860&63\\
15H121-151	&	1.21	&	1.51	&	15H		&1.17524&0.247&	1198	&7822&84\\
125H121-151	&	1.21	&	1.51	&	125H	&1.17524&0.247&	856		&7323&79\\
H121-151	&	1.21	&	1.51	&	H		&1.17524&0.247&	604		&6823&73\\
HB121-151	&	1.21	&	1.51	&	HB		&1.17524&0.247&	422		&6324&68\\	
B121-151	&	1.21	&	1.51	&	B		&1.17524&0.247&	290		&5991&64\\
15H116-158	&	1.16	&	1.58	&	15H		&1.17524&0.244&	1185	&7989&86\\
125H116-158	&	1.16	&	1.58	&	125H	&1.17524&0.244&	848		&7490&80\\
H116-158	&	1.16	&	1.58	&	H		&1.17524&0.244&	601		&6991&75\\
HB116-158	&	1.16	&	1.58	&	HB		&1.17524&0.244&	421		&6491&70\\	
B116-158	&	1.16	&	1.58	&	B		&1.17524&0.244&	291		&5992&64\\
15H125-125	&	1.25	&	1.25	&	15H		&1.08819&0.25&	1875	&7822&84\\
125H125-125	&	1.25	&	1.25	&	125H	&1.08819&0.25&	1352	&7323&79\\
H125-125	&	1.25	&	1.25	&	H		&1.08819&0.25&	966		&6823&73\\
HB125-125	&	1.25	&	1.25	&	HB		&1.08819&0.25&	683		&6324&68\\
B125-125	&	1.25	&	1.25	&	B		&1.08819&0.25&	476		&5991&64\\
\hline\hline
\end{tabular}\label{tb:model}
\end{table*}

\begin{table*}
\centering
\caption{Equations of state employed, the radius, $R_M$, and the dimensionless tidal deformability, $\Lambda_M$, for spherical neutron stars of $M=1.16$, $1.21$, $1.25$, $1.35$, $1.51$ and $1.58\,M_\odot$. $R_M$ is listed in units of ${\rm km}$.}
\begin{tabular}{c|cccccccccccc}\hline\hline
EOS&$~R_{1.16}$&$~R_{1.21}$&$~R_{1.25}$&$~R_{1.35}$&$~R_{1.51}$&$~R_{1.58}$&$~\Lambda_{1.16}$&$~\Lambda_{1.21}$&$~\Lambda_{1.25}$&$~\Lambda_{1.35}$&$~\Lambda_{1.51}$&$~\Lambda_{1.58}$\\\hline
15H 		&	13.60&	13.63&	13.65&	13.69&	13.73&	13.73&	2863&	2238&	1875&	1211&	625&	465\\
125H 	&	12.90&	12.93&	12.94&	12.97&	12.98&	12.98&	2085&	1621&	1352&	863&	435&	319\\
H 		&	12.23&	12.25&	12.26&	12.27&	12.26&	12.25&	1506&	1163&	966&	607&	298&	215\\
HB 		&	11.59&	11.60&	11.61&	11.61&	11.57&	11.53&	1079&	827&	683&	422&	200&	142\\
B 		&	11.98&	10.98&	10.98&	10.96&	10.89&	10.84&	765&	581&	476&	289&	131&	91\\\hline
\end{tabular}\label{tb:eos_model}
\end{table*}

 	The hybrid waveforms employed for deriving and calibrating our waveform model in this paper are composed of the high-frequency part ($\gtrsim 400\,{\rm Hz}$) and the low-frequency part ($\lesssim 400\,{\rm Hz}$). For the high-frequency parts, we employ our latest numerical-relativity waveforms derived partly in Ref.~\cite{Kiuchi:2017pte}. The simulations are performed by using a numerical-relativity code, {\tt SACRA}, in which an adaptive-mesh-refinement (AMR) algorithm is implemented (see Refs.~\cite{Kiuchi:2017pte} and~\cite{Yamamoto:2008js} for details of the computational setup). Binary neutron stars in quasi-circular orbits with small eccentricity $\sim10^{-3}$ are numerically derived for the initial conditions of the simulations using a spectral-method library, {\tt LORENE}~\cite{lorene}, and an eccentricity-reduction procedure described in Ref.~\cite{Kyutoku:2014yba}. 
 	
 	We employ the numerical-relativity waveforms of binary neutron stars with $m_0\approx2.7\,M_\odot$ and $m_0=2.5\,M_\odot$, where $m_0$ is the total mass of the binary at infinite separation. More precisely, equal-mass models with each mass $m_1=m_2=1.35\,M_\odot$ and $1.25\,M_\odot$, and unequal-mass models with each mass $(m_1,m_2)\approx(1.21,1.51)\,M_\odot$ and $(1.16,1.58)\,M_\odot$ are employed. We note that, for the models with each mass $(m_1,m_2)\approx(1.21,1.51)\,M_\odot$, we employ the results of the simulations of which grid resolutions are improved from those presented in Ref.~\cite{Kiuchi:2017pte}. The simulations for the new models are performed in the same way as in Ref.~\cite{Kiuchi:2017pte}. The orbital angular velocity of the initial configuration, $\Omega_0$, is chosen to be $m_0\Omega_0\approx0.0155$ and $0.0150$ for $m_0\approx2.7\,M_\odot$ and $m_0=2.5\,M_\odot$, respectively. Model parameters and grid configurations are summarized in Table~\ref{tb:model}. We note that the numerical-relativity waveforms are expected to have a phase error by $0.2$--$0.6\,{\rm rad}$ up to the time of peak amplitude (see Ref.~\cite{Kiuchi:2017pte} and Appendix~\ref{appC} for details of this estimation).
 	
 	Five parameterized piecewise-polytropic equations of state with two pieces~\cite{Read:2009yp,Lackey:2011vz,Read:2013zra,Kiuchi:2017pte} are employed to consider the cases for a wide range of binary tidal deformability, $300\lesssim{\tilde \Lambda}\lesssim1900$. For any equations of state employed in this paper, the maximum mass of spherical neutron stars is larger than $2.0\,M_\odot$, which is the approximate maximum mass among the observed neutron stars to date~\cite{Demorest:2010bx,Antoniadis:2013pzd}. The radius and the dimensionless tidal deformability of spherical neutron stars of $1.16$, $1.21$, $1.25$, $1.35$, $1.51$, and $1.58\,M_\odot$ are listed in Table~\ref{tb:eos_model}. The 15H equation of state might be incompatible with the observational results of GW170817~\cite{Abbott2017}, because the tidal deformability in this equations of state for the neutron stars of mass 1.35--$1.40,M_\odot$ is larger than 1000. However, the other equations of state are compatible with the latest observational results.
 	
 	For the low-frequency part, we employ the TEOB waveforms of Refs.~\cite{Hinderer:2016eia,Steinhoff:2016rfi,Dietrich:2017feu}, which are currently among the most successful approximants in which the tidal effects as well as higher PN effects are taken into account. There exist two types of the TEOB formalism depending on the choice of point-particle baseline; the  SEOBNRv2T and  SEOBNRv4T formalisms of which the point-particle parts agree with the SEOBNRv2 and SEOBNRv4 formalisms~\cite{Bohe:2016gbl}, respectively. In this work, we employ the  SEOBNRv2T waveforms for the low-frequency part of the hybrid waveforms. This is because the point-particle baseline of the SEOBNRv2T formalism, i.e., the SEOBNRv2 formalism, is more suitable for deriving waveforms for a non-spinning equal-mass binary (see Appendix~\ref{appA}). 
 	
 	 For each binary neutron star model in Table~\ref{tb:model}, the SEOBNRv2T waveforms are generated by specifying the mass and dimensionless tidal deformability, $\Lambda_i$ $(i=1,2)$, of each neutron star. Other tidal parameters required for generating the SEOBNRv2T waveforms, such as the octupolar tidal deformability and f-mode frequency of neutron stars, are determined from given values of $\Lambda_i$ by employing universal relations derived in Refs.~\cite{Yagi:2013sva,Chan:2014kua}. The initial gravitational-wave frequency of the SEOBNRv2T waveforms is always set to be $9\,{\rm Hz}$, and we use the spectral data only for $f\ge10\,{\rm Hz}$ to suppress the unphysical modulation due to the truncation of the waveforms at the initial time.
 	
	The hybridization of the waveforms is performed by the procedure described in Ref.~\cite{Hotokezaka:2016bzh}. First, we align the time and phase of the SEOBNRv2T waveforms and the numerical-relativity waveforms by searching for $t_{\rm s}$'s and $\phi_{\rm s}$'s that minimize
\begin{align}
	I=\int^{t_{\rm max}}_{t_{\rm min}} \left|h_{\rm NR}\left(t_{\rm ret}'\right)-h_{\rm TEOB}\left(t_{\rm ret}'+t_{\rm s}\right) e^{i\phi_{\rm s}}\right|^2dt_{\rm ret}',
 \end{align}
where $t_{\rm ret}$ is the retarded time of the simulation, $h_{\rm NR}$ and $h_{\rm TEOB}$ are the time-domain complex waveforms derived by numerical-relativity simulation and the SEOBNRv2T formalism, respectively. Here, the complex waveform, $h$, is defined by $h=h_+-ih_\times$, with $h_+$ and $h_\times$ denoting the plus and cross modes of gravitational waves, respectively. We choose $t_{\rm min}=20\,{\rm ms}$ and $t_{\rm max}=40\,{\rm ms}$ following Ref.~\cite{Kiuchi:2017pte}. After the alignment, two waveforms are hybridized as
 \begin{align}
 	&h_{\rm Hybrid}\left(t\right)=\nonumber\\&
 	\left\{
 	\begin{array}{cc}
 		h_{\rm TEOB}\left(t\right)	&	t\le t_{\rm min},\\
 		\left[1-H\left(t\right)\right]h_{\rm TEOB}\left(t\right)+H\left(t\right)h_{\rm NR}\left(t\right)	&	t_{\rm min}<t<t_{\rm max},\\
 		h_{\rm NR}\left(t\right)	&	t_{\rm max}\le t,
 	\end{array}
 	\right.
\end{align}
where we choose a Hann window function for $H\left(t\right)$ as
\begin{align}
	H\left(t\right)=\frac{1}{2}\left[1-\cos\left(\pi\frac{t-t_{\rm min}}{t_{\rm max}-t_{\rm min}}\right)\right].
\end{align}
	We find that the hybrid waveforms depend only weakly on the choices of $t_{\rm min}$ and $t_{\rm max}$. For example, employing $t_{\rm min}=25\,{\rm ms}$ and $t_{\rm max}=45\,{\rm ms}$ instead changes the phase of the hybrid waveforms only by $\lesssim0.1\,{\rm rad}$ up to the time of the peak amplitude, and in particular, the change in the phase is always smaller than $0.05\,{\rm rad}$ until the gravitational-wave frequency reaches $1000\,{\rm Hz}$.
	
\subsection{Computing the Fourier spectrum}\label{sec2b}
The Fourier spectrum of gravitational waves, ${\tilde h}\left(f\right)$, is defined by~\cite{Cutler:1994ys}
\begin{align}
	{\tilde h}\left(f\right)=\int_{t_{\rm i}}^{t_{\rm f}} h_+\left(t\right){\rm e}^{-2\pi i ft} dt,
\end{align}
where $t_{\rm i}$ and $t_{\rm f}$ are the initial and final time of the waveform data, respectively. Note that, for binary neutron stars, the Fourier transformation of $h_\times$ results approximately in $-i{\tilde h}$. 

To suppress the unphysical modulation in the spectrum, we adopt a window function, $w\left(t\right)$, in the initial and final time of the waveform data. We employ a tapered cosine filter for $w\left(t\right)$ which is defined by  
\begin{align}
	w&\left(t\right)=\nonumber\\
	&\left\{
	\begin{array}{cc}
		\left\{1-{\rm cos}\left[\pi\left(t-t_{\rm i}\right)/\Delta t_{\rm i}\right]\right\}/2 &t_{\rm i}\le t<t_{\rm i}+\Delta t_{\rm i},\\
		1 & t_{\rm i}+\Delta t_{\rm i}\le t<t_{\rm f}-\Delta t_{\rm f},\\
		\left\{1-{\rm cos}\left[\pi\left(t_{\rm f}-t\right)/\Delta t_{\rm f}\right]\right\}/2 & t_{\rm f}-\Delta t_{\rm f}\le t<t_{\rm f},
	\end{array}
	\right.
\end{align}
where $\Delta t_{\rm i}$ and $\Delta t_{\rm f}$ are the widths of the tapering regions. We choose $\Delta t_{\rm i}\approx 10\,{\rm s}$ and $\Delta t_{\rm f}=100\,m_{\rm 0}$.

The amplitude of the spectrum can be obtained directly from the absolute value of ${\tilde h}\left(f\right)$. To obtain $\Psi\left(f\right)$ as a continuous function of $f$, we integrate $d\Psi/df\left(f\right)$ in frequency as
\begin{align}
	\Psi\left(f\right)=\int^f \frac{d\Psi}{df}\left(f'\right)df',
\end{align}
where $d\Psi/df\left(f\right)$ is calculated by
\begin{align}
	\frac{d\Psi}{df}\left(f\right)=-\frac{1}{\left|{\tilde h}\left(f\right)\right|^2}{\rm Im}\left[{\tilde h}^*\left(f\right)\frac{ d{\tilde h}}{df}\left(f\right)\right],
\end{align}
and ${\tilde h}^*\left(f\right)$ is the complex conjugate of ${\tilde h}\left(f\right)$.

	$\Psi\left(f\right)$ has degrees of freedom to shift its value by
\begin{align}
	\Psi\left(f\right)\rightarrow\Psi\left(f\right)+2\pi f t_0-\phi_0,
\end{align}
where $t_0$ and $\phi_0$ can be chosen arbitrarily. Thus, to compare the phases of different waveforms, we need to align the time and phase origins of each phase. For this purpose, we define the difference between gravitational-wave phases, $\Psi_1$ and $\Psi_2$, by 
\begin{align}
	\Delta\Psi(f)=\Psi_1(f)-\Psi_2(f)-2\pi f t_0+\phi_0, \label{eq:align}
\end{align}
where $t_0$ and $\phi_0$ are determined by minimizing
\begin{align}
	I'=\int^{f_{\rm 1}}_{f_{\rm 0}} \left|\Psi_1\left(f\right)-\Psi_2\left(f\right)-2\pi f t_0+\phi_0\right|^2df,\label{eq:dtdphi}
\end{align}
and $f_0$ and $f_1$ are the lower-bound and upper-bound frequencies of the alignment, respectively. We note that, in the following, we always align the phases by this procedure to plot the phase difference. 
\subsection{Tidal part model for the gravitational-wave phase}\label{sec2c}
\subsubsection{Equal-mass cases}\label{sec2c1}
\begin{figure}
 	 \includegraphics[width=1\linewidth]{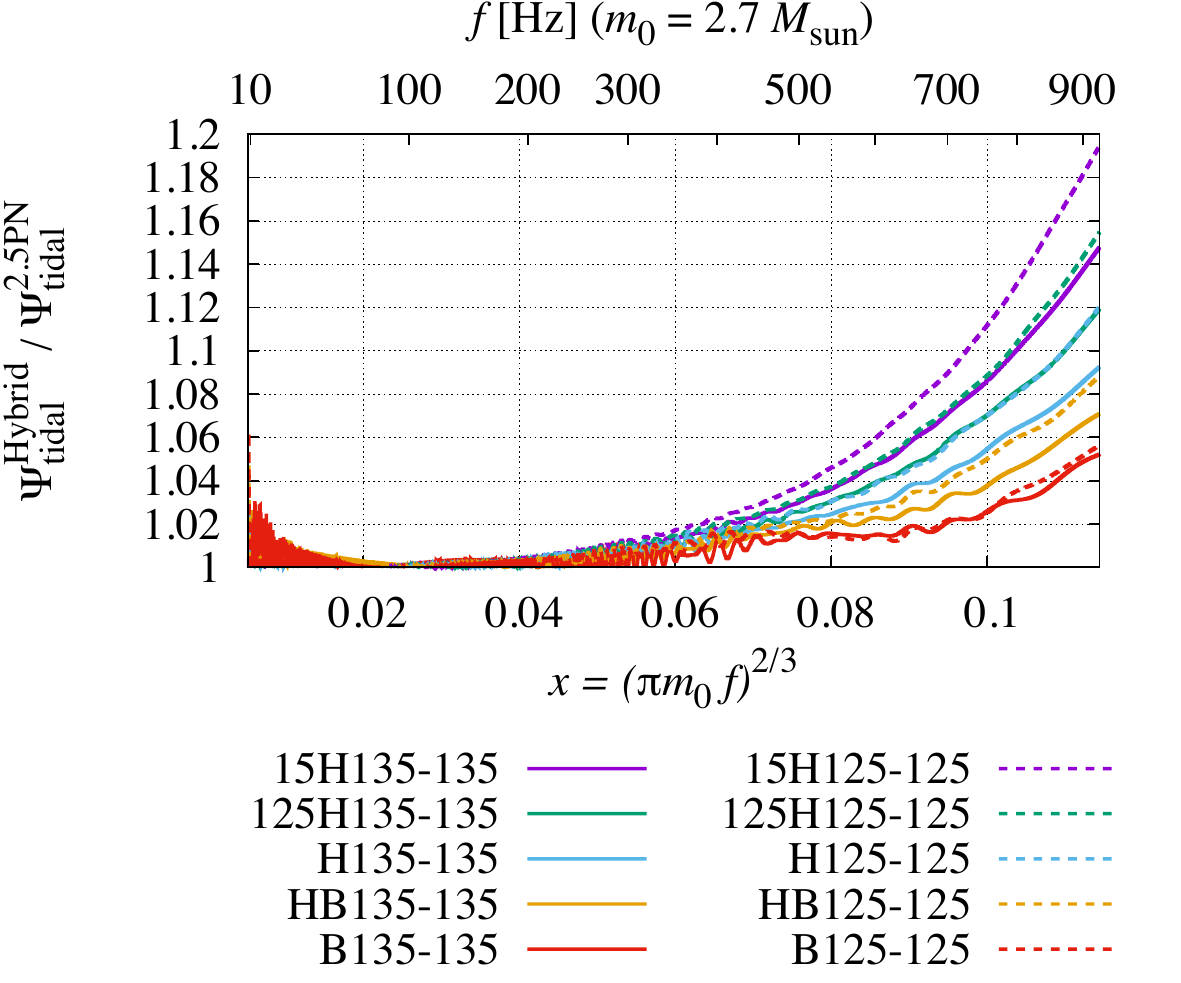}
 	 \caption{The tidal part of the gravitational-wave phase computed from the hybrid waveforms for the equal-mass binaries, $\Psi_{\rm tidal}^{\rm Hybrid}$, normalized by the 2.5 PN order tidal-part phase, $\Psi_{\rm tidal}^{\rm 2.5PN}$, given by Eq.~\eqref{eq:tidal25pn}. The horizontal axis denotes a dimensionless PN parameter of the binary. Each tidal-part phase of the hybrid waveforms is aligned with the 2.5 PN order tidal-part phase, $\Psi^{\rm 2.5PN}_{\rm tidal}$, for $10\,{\rm Hz}\le f \le 50\,{\rm Hz}$ employing Eqs.~\eqref{eq:align} and~\eqref{eq:dtdphi}.}\label{fig:H}
\end{figure}
	First, we derive a phase model for the hybrid waveforms focusing on equal-mass cases. Figure~\ref{fig:H} shows the tidal part of the gravitational-wave phase computed from the hybrid waveforms, $\Psi_{\rm tidal}^{\rm Hybrid}$ ($=\Psi^{\rm Hybrid}-\Psi_{\rm pp}$, where $\Psi^{\rm Hybrid}$ is the phase of the hybrid waveforms), for the equal-mass binaries normalized by the 2.5 PN order (equal-mass) tidal-part phase given by\footnote{Strictly speaking, this formula is not complete up to the 2.5 PN order because the 2 PN order tidal correction to gravitational-radiation reaction is neglected. We overlook such correction in this work because it is expected to be sub-dominant~\cite{Damour:2012yf}.}~\cite{Damour:2012yf}
\begin{align}
&\Psi_{\rm tidal}^{\rm 2.5PN}=\frac{3}{32}\left(-\frac{39}{2}{\Lambda}\right)x^{5/2}\nonumber\\
&\times\left[1+\frac{3115}{1248}x-\pi x^{3/2}+\frac{28024205}{3302208} x^2 -\frac{4283}{1092}\pi x^{5/2}\right],\label{eq:tidal25pn}
\end{align}
where $x=\left(\pi m_0 f\right)^{2/3}$ is a dimensionless PN parameter, and $\Lambda=\Lambda_1=\Lambda_2$ for the equal-mass cases. We note that the tidal-part phase of the hybrid waveforms in Fig.~\ref{fig:H} is aligned with the 2.5 PN order tidal-part phase given by Eq.~\eqref{eq:tidal25pn} for $10\,{\rm Hz}\le f \le 50\,{\rm Hz}$ employing Eqs.~\eqref{eq:align} and~\eqref{eq:dtdphi}. We find that the tidal-part phase of the hybrid waveforms deviates significantly from the 2.5 PN order tidal-part phase in the high-frequency range, $f\gtrsim500\,{\rm Hz}$ ($x\gtrsim0.075$ for $m_0=2.7\,M_\odot$), and the deviation depends non-linearly on $\Lambda$ (note the quantities shown in Fig.~\ref{fig:H} are already normalized by $\Lambda$). This indicates that the non-linear contribution of $\Lambda$ is appreciably present in $\Psi_{\rm tidal}^{\rm Hybrid}$ for the high-frequency range.

\begin{figure}
 	 \includegraphics[width=1\linewidth]{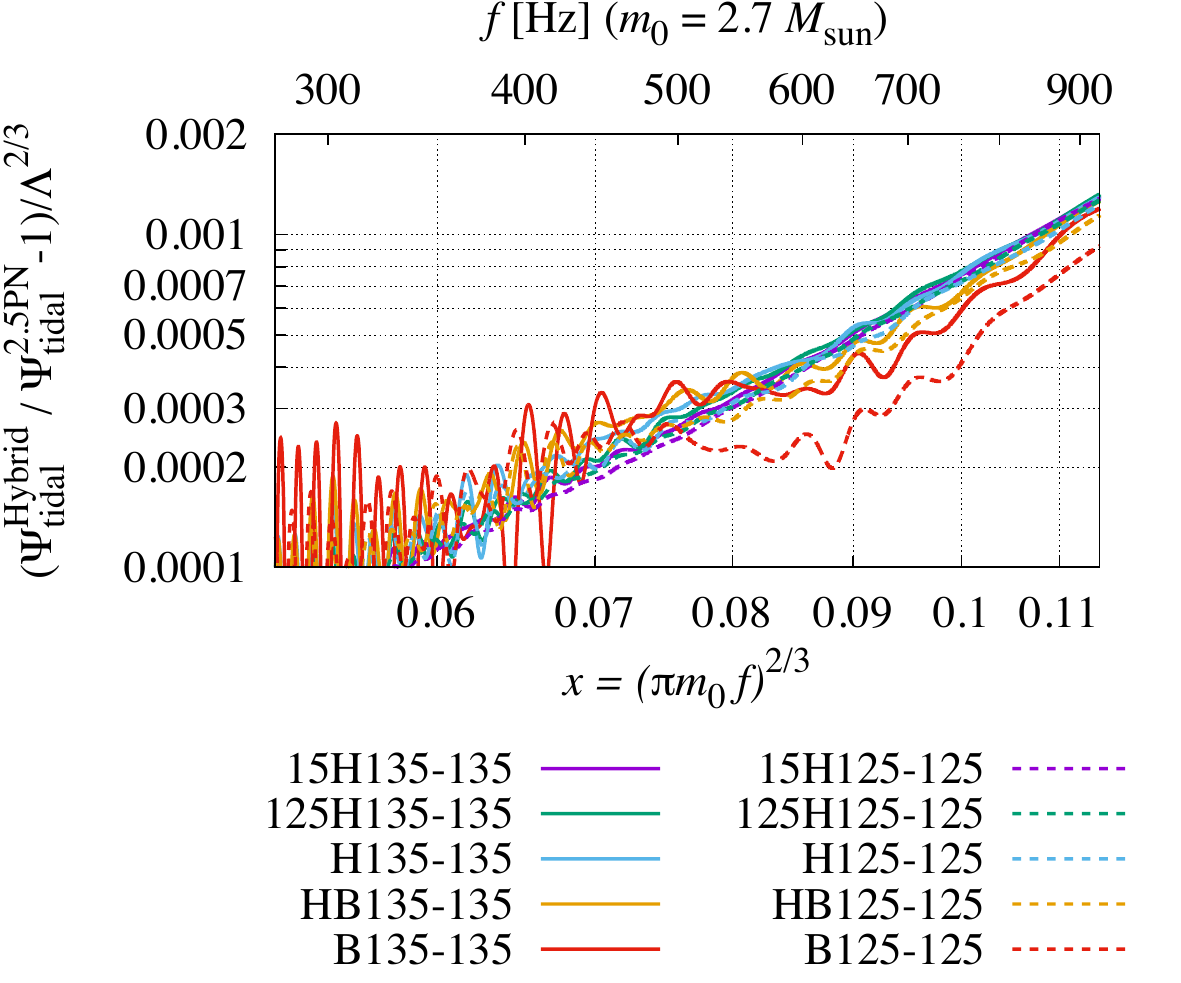}
 	 \caption{The relative deviation of the tidal-part phase of the hybrid waveforms from the 2.5 PN order tidal-part phase normalized by $\Lambda^{2/3}$, i.e., $(\Psi_{\rm tidal}^{\rm Hybrid}/\Psi_{\rm tidal}^{\rm 2.5PN}-1)/\Lambda^{2/3}$. Both vertical and horizontal axes are plotted with logarithmic scales.}\label{fig:H-log}
\end{figure}
In Fig.~\ref{fig:H-log}, we plot the relative deviation of the tidal-part phase of the hybrid waveforms from the 2.5 PN order tidal-part phase normalized by $\Lambda^{2/3}$, i.e., $(\Psi_{\rm tidal}^{\rm Hybrid}/\Psi_{\rm tidal}^{\rm 2.5PN}-1)/\Lambda^{2/3}$. Figure~\ref{fig:H-log} clearly shows that the relative deviation can be well approximated by a power law in $x$. Furthermore, it shows that the relative deviation is approximately proportional to $\Lambda^{2/3}$ because all the curves align. We note that, exceptionally, the relative deviation for the B equation of state shows a slightly different trend from the other cases. The reason for this is that the tidal deformability is so small that its effect cannot be accurately extracted from the numerical-relativity waveform for such a soft equation of state.
	
	To correct this deviation, we extend the 2.5 PN order tidal-part phase formula of Eq.~\eqref{eq:tidal25pn} by multiplying a non-linear correction to $\Lambda$ as
\begin{align}
&\Psi_{\rm tidal}^{\rm em}=\frac{3}{32}\left[-\frac{39}{2}{\Lambda}\left(1+a{\Lambda}^{2/3} x^p\right)\right]x^{5/2}\nonumber\\
&\times\left[1+\frac{3115}{1248}x-\pi x^{3/2}+\frac{28024205}{3302208} x^2 -\frac{4283}{1092}\pi x^{5/2}\right],\label{eq:phiT_H_eq}
\end{align}
where $a$ and $p$ are fitting parameters. We note that the exponent of the nonlinear term in $\Lambda$, $p$, is deduced to be $\approx 2/3$ even if it is also set to be a fitting parameter and determined by employing several hybrid waveforms. The fitting parameters, $a$ and $p$, are determined by minimizing
\begin{align}
I''=\int_{f_{\rm min}}^{f_{\rm max}}\left|\Psi^{\rm Hybrid}_{\rm tidal}\left(f\right)-\Psi_{\rm tidal}^{\rm em}\left(f\right)-2\pi f t_0+\phi_0\right|^2df,\label{eq:fitTidal}
\end{align}
where $t_0$ and $\phi_0$ are parameters that correspond to the degrees of freedom for choosing the time and phase origins. Thus, we minimize $I''$ for the four parameters, $a$, $p$, $t_0$, and $\phi_0$. The fitting is performed for $f_{\rm min}=10$\,Hz and $f_{\rm max}=1000$\,Hz.

\begin{figure}
 	 \includegraphics[width=1\linewidth]{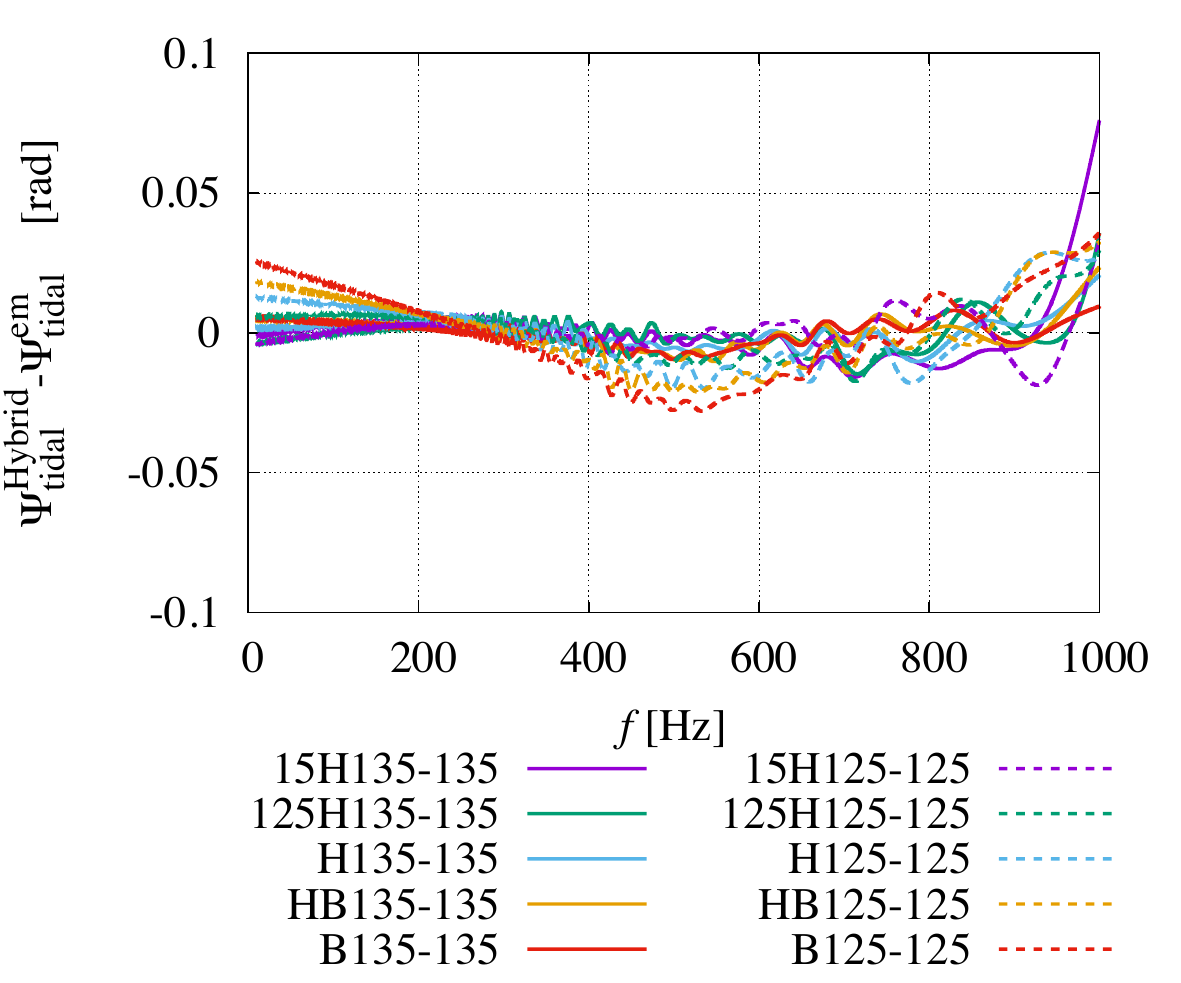}
 	 \caption{The difference between the tidal-part phase of the hybrid waveforms and the tidal-part phase model described in Eq.~\eqref{eq:phiT_H_eq} for the equal-mass cases, where the phases are aligned for $10\,{\rm Hz}\le f\le1000\,{\rm Hz}$ employing Eqs.~\eqref{eq:align} and~\eqref{eq:dtdphi}.}\label{fig:Hcomp}
\end{figure}
We use the hybrid waveform of 15H125-125 for determining the fitting parameters because the non-linear contribution of $\Lambda$ is most significant for this among the binary neutron star models employed in this work. Then we obtain
\begin{align}
	a&=12.55,\nonumber\\
	p&=4.240.\label{eq:best_fits}
\end{align}
Figure~\ref{fig:Hcomp} shows the phase difference between the tidal-part of the hybrid waveforms and the tidal-part phase model of Eq.~\eqref{eq:phiT_H_eq} for the equal-mass cases, where two phases are aligned for $10\,{\rm Hz}\le f\le1000\,{\rm Hz}$ employing Eqs.~\eqref{eq:align} and~\eqref{eq:dtdphi}. Although the fitting parameters are determined by employing only the hybrid waveform of 15H125-125 as a reference, we find that the error in the tidal-part phase model, Eq.~\eqref{eq:phiT_H_eq}, is always smaller than $0.05\,{\rm rad}$ except for 15H135-135. This result indicates that there is only a small difference between the waveform models determined from different hybrid waveforms (see Appendix~\ref{appD}). The phase error for 15H135-135 is as large as $0.08\,{\rm rad}$ for $f\approx 1000\,{\rm Hz}$. However, it is smaller than the phase error in the numerical-relativity waveforms associated with the finite-differencing~\cite{Kiuchi:2017pte}.

\subsubsection{Unequal-mass cases}\label{sec2c2}
Next, we extend the tidal-part phase model of Eq.~\eqref{eq:phiT_H_eq} to unequal-mass cases. Considering the dependence on the symmetric mass ratio, the 1 PN order tidal correction to the phase can be written in terms of the symmetric and anti-symmetric contributions of neutron-star tidal deformation as~\cite{Wade:2014vqa}
\begin{align}
\Psi_{\rm tidal}^{1{\rm PN}}&=\frac{3}{128\eta}\left(-\frac{39}{2}{\tilde \Lambda}\right)x^{5/2}\nonumber\\
&\times\left[1+\left(\frac{3115}{1248}-\frac{6595}{7098}\sqrt{1-4\eta}\frac{\delta{\tilde \Lambda}}{{\tilde \Lambda}}\right)x\right],
\end{align}
where ${\tilde \Lambda}$ and $\delta{\tilde \Lambda}$ are defined by
\begin{align}
{\tilde \Lambda}&=\frac{8}{13}\left[\left(1+7\eta-31\eta^2\right)\left(\Lambda_1+\Lambda_2\right)\right.\nonumber
\\&-\left.\sqrt{1-4\eta}\left(1+9\eta-11\eta^2\right)\left(\Lambda_1-\Lambda_2\right)\right]
\label{eq:symlambda}
\end{align}
and
\begin{align}
\delta{\tilde \Lambda}&=\frac{1}{2}\left[\sqrt{1-4\eta}\left(1-\frac{13272}{1319}\eta+\frac{8944}{1319}\eta^2\right)\left(\Lambda_1+\Lambda_2\right)\right.\nonumber
\\&-\left.\left(1-\frac{15910}{1319}\eta+\frac{32850}{1319}\eta^2+\frac{3380}{1319}\eta^3\right)\left(\Lambda_1-\Lambda_2\right)\right],
\end{align}
respectively. We refer to ${\tilde \Lambda}$ as the binary tidal deformability. For realistic cases, the tidal contributions to the gravitational-wave phase are dominated by the contributions from the ${\tilde \Lambda}$ terms~\cite{Wade:2014vqa}. Assuming that the contributions from the $\delta{\tilde \Lambda}$ terms and those from the higher-order terms are always sub-dominant in the tidal part of the phase, we extend the formula of Eq.~\eqref{eq:phiT_H_eq} by replacing $3/32$ to $3/128\eta$~\cite{Khan:2015jqa} and $\Lambda$ to ${\tilde \Lambda}$ as
\begin{align}
&\Psi_{\rm tidal}=\frac{3}{128\eta}\left[-\frac{39}{2}{\tilde \Lambda}\left(1+a{\tilde \Lambda}^{2/3} x^p\right)\right]x^{5/2}\nonumber
\\&\times\left[1+\frac{3115}{1248}x-\pi x^{3/2}+\frac{28024205}{3302208} x^2 -\frac{4283}{1092}\pi x^{5/2}\right],\label{eq:phiT_H}
\end{align}
where the values in Eq.~\eqref{eq:best_fits} are used for $a$ and $p$.

\begin{figure}
 	 \includegraphics[width=1\linewidth]{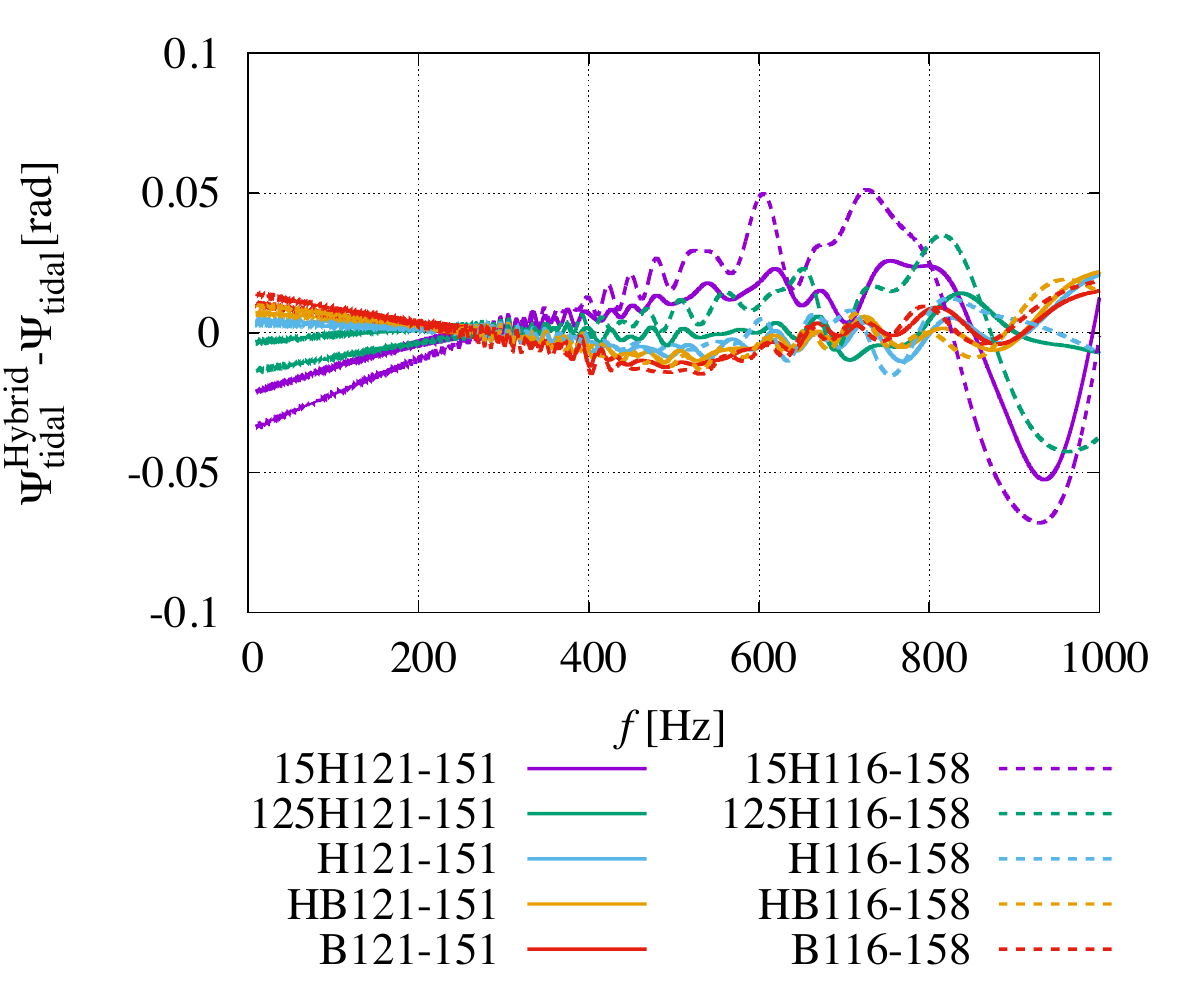}
 	 \caption{The difference between the tidal-part phase of the hybrid waveforms and the tidal-part phase model described in Eq.~\eqref{eq:phiT_H} for the unequal-mass cases. Here, the phases are aligned for $10\,{\rm Hz}\le f\le1000\,{\rm Hz}$ employing Eqs.~\eqref{eq:align} and~\eqref{eq:dtdphi}.}\label{fig:H_ne}
\end{figure}
Figure~\ref{fig:H_ne} shows the phase difference between the hybrid waveforms and the tidal-part phase model described in Eq.~\eqref{eq:phiT_H} for the unequal-mass cases. Here, two phases are again aligned for $10\,{\rm Hz}\le f\le1000\,{\rm Hz}$ employing Eqs.~\eqref{eq:align} and~\eqref{eq:dtdphi}. Although the fitting parameters are determined only by employing the hybrid waveform of 15H125-125, Eq.~\eqref{eq:best_fits}, we find that the phase error is always smaller than $\approx0.07\,{\rm rad}$ for these unequal-mass cases. 
 
\subsection{Tidal part model for the gravitational-wave amplitude}\label{sec2d}
 \begin{figure}
 	 \includegraphics[width=1\linewidth]{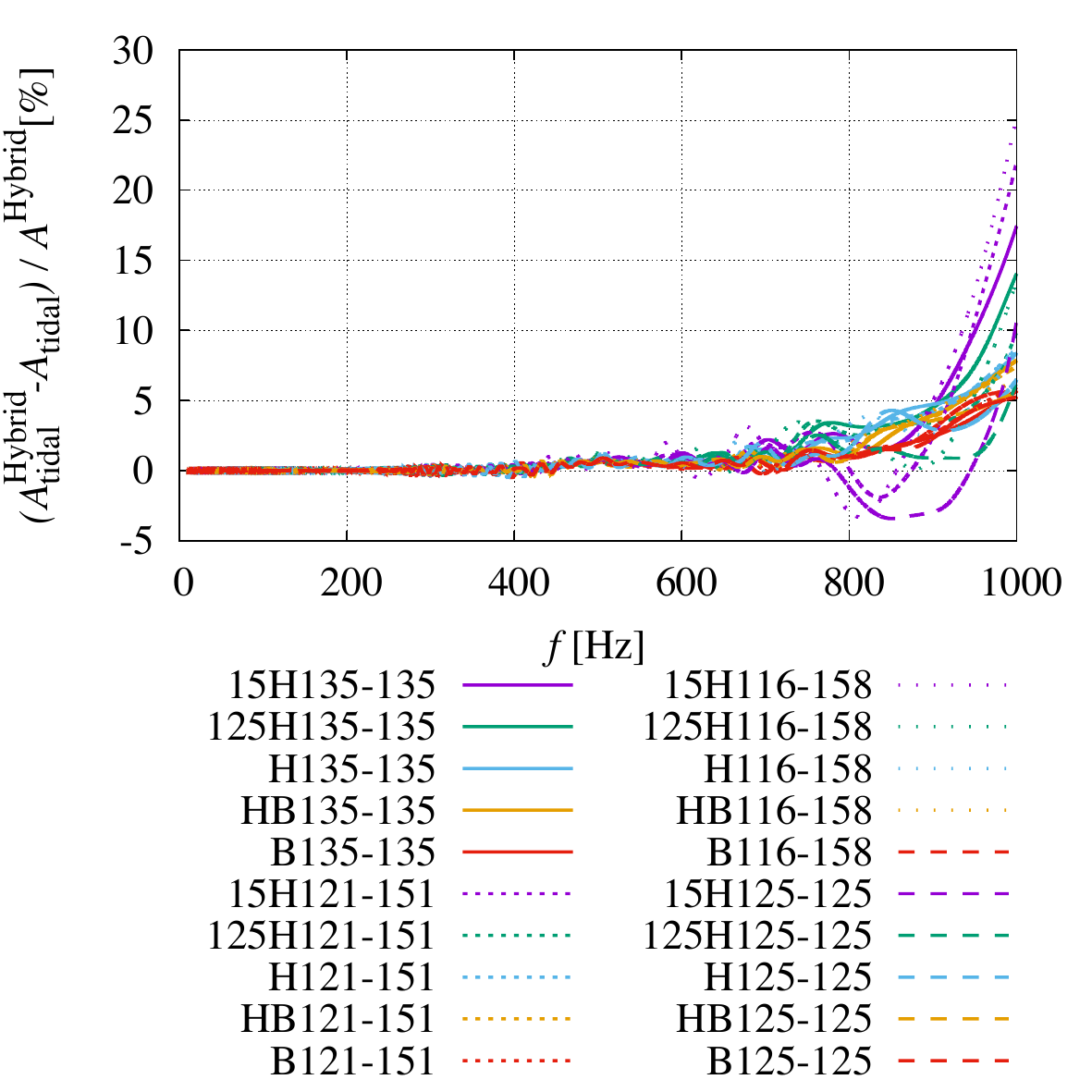}
 	 \caption{The difference between the tidal-part amplitude given by Eq.~\eqref{eq:ampH_ne}, $A_{\rm tidal}$, and that of the hybrid waveforms, $A_{\rm tidal}^{\rm Hybrid}$, normalized by the amplitude of the hybrid waveforms, $A^{\rm Hybrid}$.}\label{fig:amp}
\end{figure} 
We derive the tidal-part amplitude model in the same approach as we took for the phase model: First, we derive the tidal-part amplitude model for the hybrid waveforms for equal-mass cases, and then we extend it to unequal-mass cases. 

The tidal-part amplitude model for the hybrid waveforms is derived based on the 1 PN order (equal-mass) formula for the tidal-part amplitude given by~\cite{Vines:2011ud,Damour:2012yf,Hotokezaka:2016bzh}
\begin{align}
	A_{\rm tidal}^{\rm 1PN}=\sqrt{\frac{5\pi}{96}}\frac{m_0^2}{D_{\rm eff}} \Lambda x^{-7/4}\left(-\frac{27}{16}x^{5}-\frac{449}{64}x^{6}\right),\label{eq:amp1PN}
\end{align}
where $D_{\rm eff}$ is the effective distance to the binary (see Ref.~\cite{Hotokezaka:2016bzh} for its definition). To take the higher-order PN tidal effects into account, we add a polynomial term to Eq.~\eqref{eq:amp1PN} as
\begin{align}
	A_{\rm tidal}^{\rm em}=\sqrt{\frac{5\pi}{96}}\frac{m_0^2}{D_{\rm eff}} \Lambda x^{-7/4}\left(-\frac{27}{16}x^{5}-\frac{449}{64}x^{6}+b x^q\right),\label{eq:ampH}
\end{align}
where $b$ and $q$ are the fitting parameters. We determine $b$ and $q$ by minimizing
\begin{align}
I'''=\int_{f_{\rm min}}^{f_{\rm max}}\left|A^{\rm Hybrid}_{\rm tidal}\left(f\right)-A^{\rm em}_{\rm tidal}\left(f\right)\right|^2df,\label{eq:fitamp}
\end{align}
where $A^{\rm Hybrid}_{\rm tidal}$ is the tidal-part amplitude of the hybrid waveforms, $f_{\rm min}$ and $f_{\rm max}$ are set to be $10\,{\rm Hz}$ and $1000\,{\rm Hz}$, respectively. Employing the hybrid waveform of 15H125-125 as a reference, we obtain $b=-4251$ and $q=7.890$.

As in the phase model, we extend Eq.~\eqref{eq:ampH} to unequal-mass cases by replacing the leading order coefficient, $\sqrt{5\pi/96}$, and $\Lambda$ to $\sqrt{5\pi\eta/24}$ and ${\tilde \Lambda}$, respectively, as
\begin{align}
	A_{\rm tidal}=\sqrt{\frac{5\pi\eta}{24}}\frac{m_0^2}{D_{\rm eff}} {\tilde \Lambda} x^{-7/4}\left(-\frac{27}{16}x^{5}-\frac{449}{64}x^{6}+b x^q\right).\label{eq:ampH_ne}
\end{align}
Figure~\ref{fig:amp} shows the relative error of the tidal-part amplitude model defined by $(A_{\rm tidal}^{\rm Hybrid}-A_{\rm tidal})/A^{\rm Hybrid}$, where $A^{\rm Hybrid}$ is the amplitude of the hybrid waveforms. For ${\tilde \Lambda}\le850$, the relative error of the tidal-part amplitude model is always smaller than $10\%$. The relative error is larger for ${\tilde \Lambda}\ge850$, and in particular, it is larger than $15\%$ for 15H135-135, 15H121-151, and 15H116-158. However, such large values of the error are only present for $f\gtrsim900\,{\rm Hz}$, and they have only minor effects on the accuracy of our waveform model as is shown in the next section.
	
\section{Validity of the analytic model}\label{sec3}
 	We constructed a frequency-domain gravitational-waveform model for binary neutron stars by employing the tidal-part and point-particle part models of gravitational waves derived in the previous section and Appendix~\ref{appA}, respectively, as
\begin{align}
	{\tilde h}_{\rm model}&={\tilde h}_{\rm model}\left(f;{\cal M}_{\rm c},\eta,{\tilde \Lambda},\phi_0,t_0,D_{\rm eff}\right)\\\nonumber
	&=\left(A_{\rm TF2+}+A_{\rm tidal}\right){\rm e}^{-i\left(\Psi_{\rm TF2+}+\Psi_{\rm tidal}\right)}.
\end{align}
This waveform model has 6 parameters, $\left\{\theta_i\right\}_{i=1}^6=\left\{{\cal M}_{\rm c},\eta,{\tilde \Lambda},\phi_0,t_0,D_{\rm eff}\right\}$. In this section we check the validity of our waveform model using the hybrid waveforms as hypothetical signals.

\subsection{Distinguishability}
	To check the validity of our waveform model derived in the previous section, we calculate the distinguishability between our waveform model and the hybrid waveforms supposing advanced LIGO as a fiducial detector. For this purpose, we define an inner product and the norm of the waveforms by
\begin{align}
	\left({\tilde h}_1\middle|{\tilde h}_2\right)=4{\rm Re}\left[\int_{f_{\rm min}}^{f_{\rm max}}
	\frac{{\tilde h}_1\left(f\right){\tilde h}^*_2\left(f\right)}{S_{\rm n}\left(f\right)}df\right],\label{eq:inp}
\end{align}
and
\begin{align}
	\rho=||{\tilde h}||=\sqrt{\left({\tilde h}\middle|{\tilde h}\right)},
\end{align}
respectively, where $S_{\rm n}$ denotes the one-sided noise spectrum density of the detector. The distinguishability between two waveforms, ${\tilde h}_1$ and ${\tilde h}_2$, is defined by~\cite{Lindblom:2008cm,Read:2013zra}
\begin{align}
	\Delta\rho\left({\tilde h}_1,{\tilde h}_2\right)=\min_{\phi_0,t_0}||{\tilde h}_1-{\tilde h}_2\left(\phi_0,t_0\right)||,
\end{align}
where $\phi_0$ and $t_0$ are arbitrary phase and time shifts of the waveforms, respectively. We also define the mismatch (or unfaithfulness) between two waveforms, ${\tilde h}_1$ and ${\tilde h}_2$, by
\begin{align}
	{\bar F}=1-\max_{\phi_0,t_0}\frac{\left({\tilde h}_1\middle|{\tilde h}_2\left(\phi_0,t_0\right)\right)}{||{\tilde h}_1||\,||{\tilde h}_2||}.\label{eq:mismatch}
\end{align}
Throughout this paper, we employ the noise spectrum density of the {\tt ZERO\_DETUNED\_HIGH\_POWER} configuration of advanced LIGO~\cite{aLIGOnoise} for $S_{\rm n}$. The lower and upper bounds of the integration in Eq.~\eqref{eq:inp} are set to be $10\,{\rm Hz}$ and $1000\,{\rm Hz}$, respectively. We note that $\rho$ corresponds to the signal-to-noise ratio~\cite{Cutler:1994ys}, and $\Delta \rho=1$ indicates that two waveforms are distinguishable approximately at the $1\sigma$ level~\cite{Lindblom:2008cm}. The signal-to-noise ratio and the distinguishability are proportional to the inverse of the effective distance, $D_{\rm eff}$. 

\begin{table*}[t]
\centering
\caption{The distinguishability between our waveform model and the hybrid waveforms. The distinguishability of the SEOBNRv2T waveforms and PN waveform models with respect to the hybrid waveforms is also shown. The number in the parentheses denotes the mismatch with respect to the hybrid waveforms defined by Eq.~\eqref{eq:mismatch}. The signal-to-noise ratio is always normalized to 50. We note that the parameters of our waveform model are determined employing the hybrid waveform of 15H125-125.}
\begin{tabular}{c|c|cccc}
\hline\hline
Model		 &~~~${\tilde \Lambda}$~~~& our waveform model	&	SEOBNRv2T	&	PNtidal(TF2)&	PNtidal(TF2+)\\\hline
15H135-135	&1211&~~0.14~~($4.1\times10^{-6}$)	&~~0.54~~($5.0\times10^{-5}$)&~~3.86~~($3.0\times10^{-3}$)&~~2.68~~($1.4\times10^{-3}$)\\
125H135-135	&863&~~0.14~~($4.0\times10^{-6}$)	&~~0.25~~($1.2\times10^{-5}$)&~~3.02~~($1.8\times10^{-3}$)&~~1.67~~($5.6\times10^{-4}$)\\
H135-135	&607&~~0.12~~($2.9\times10^{-6}$)	&~~0.12~~($3.0\times10^{-6}$)&~~2.48~~($1.2\times10^{-3}$)&~~0.95~~($1.8\times10^{-4}$)\\
HB135-135	&422&~~0.11~~($2.6\times10^{-6}$)	&~~0.14~~($3.7\times10^{-6}$)&~~2.18~~($9.5\times10^{-4}$)&~~0.50~~($5.0\times10^{-5}$)\\
B135-135	&289&~~0.10~~($2.0\times10^{-6}$)	&~~0.16~~($5.3\times10^{-6}$)&~~2.04~~($8.3\times10^{-4}$)&~~0.25~~($1.3\times10^{-5}$)\\
15H121-151	&1198&~~0.18~~($6.3\times10^{-6}$)	&~~0.68~~($8.5\times10^{-5}$)&~~4.05~~($3.3\times10^{-3}$)&~~2.79~~($1.6\times10^{-3}$)\\
125H121-151	&856&~~0.11~~($2.5\times10^{-6}$)	&~~0.26~~($1.3\times10^{-5}$)&~~3.16~~($2.0\times10^{-3}$)&~~1.70~~($5.7\times10^{-4}$)\\
H121-151	&604&~~0.12~~($2.8\times10^{-6}$)	&~~0.11~~($2.3\times10^{-6}$)&~~2.62~~($1.4\times10^{-3}$)&~~0.96~~($1.8\times10^{-4}$)\\
HB121-151	&422&~~0.12~~($3.0\times10^{-6}$)	&~~0.12~~($3.1\times10^{-6}$)&~~2.32~~($1.1\times10^{-3}$)&~~0.51~~($5.2\times10^{-5}$)\\	
B121-151	&290&~~0.13~~($3.1\times10^{-6}$)	&~~0.19~~($7.2\times10^{-6}$)&~~2.16~~($8.8\times10^{-4}$)&~~0.24~~($1.1\times10^{-5}$)\\
15H116-158	&1185&~~0.24~~($1.1\times10^{-5}$)	&~~0.74~~($1.1\times10^{-4}$)&~~4.23~~($3.6\times10^{-3}$)&~~2.88~~($1.7\times10^{-3}$)\\
125H116-158	&848&~~0.14~~($3.8\times10^{-6}$)	&~~0.34~~($2.4\times10^{-5}$)&~~3.34~~($2.2\times10^{-3}$)&~~1.77~~($6.3\times10^{-4}$)\\
H116-158	&601&~~0.12~~($3.0\times10^{-6}$)	&~~0.12~~($2.8\times10^{-6}$)&~~2.76~~($1.5\times10^{-3}$)&~~0.98~~($1.9\times10^{-4}$)\\
HB116-158	&421&~~0.14~~($4.0\times10^{-6}$)	&~~0.11~~($2.4\times10^{-6}$)&~~2.45~~($1.2\times10^{-3}$)&~~0.50~~($5.1\times10^{-5}$)\\	
B116-158	&291&~~0.16~~($5.0\times10^{-6}$)	&~~0.15~~($4.6\times10^{-6}$)&~~2.28~~($1.0\times10^{-3}$)&~~0.22~~($1.0\times10^{-5}$)\\
15H125-125	&1875&~~0.09~~($1.6\times10^{-6}$) 	&~~0.83~~($1.2\times10^{-4}$)&~~4.63~~($4.3\times10^{-3}$)&~~3.51~~($2.5\times10^{-3}$)\\
125H125-125	&1352&~~0.09~~($1.6\times10^{-6}$) 	&~~0.34~~($2.4\times10^{-5}$)&~~3.61~~($2.6\times10^{-3}$)&~~2.31~~($1.1\times10^{-3}$)\\
H125-125	&966&~~0.13~~($3.5\times10^{-6}$) 	&~~0.18~~($6.6\times10^{-6}$)&~~2.83~~($1.6\times10^{-3}$)&~~1.36~~($3.7\times10^{-4}$)\\
HB125-125	&683&~~0.16~~($5.0\times10^{-6}$)	&~~0.20~~($7.8\times10^{-6}$)&~~2.36~~($1.1\times10^{-3}$)&~~0.71~~($1.0\times10^{-4}$)\\
B125-125	&476&~~0.20~~($8.3\times10^{-6}$)	&~~0.23~~($1.0\times10^{-5}$)&~~2.10~~($8.8\times10^{-4}$)&~~0.30~~($1.8\times10^{-5}$)\\
\hline\hline
\end{tabular}\label{tb:dsnr_comp}
\end{table*}

In Table~\ref{tb:dsnr_comp}, we summarize the distinguishability between our waveform model and the hybrid waveforms. Here, the signal-to-noise ratio is always fixed to be 50 by adjusting $D_{\rm eff}$ because the tidal deformability is clearly measurable only for events with a high signal-to-noise ratio. For comparison, we also compute the distinguishability of the SEOBNRv2T waveforms and PN waveform models with respect to the hybrid waveforms. For the tidal part of the PN waveform models, we employ the 2.5 PN order phase and the 1 PN order amplitude formulas given by~\cite{Vines:2011ud,Damour:2012yf,Hotokezaka:2016bzh}
\begin{align}
&\Psi_{\rm tidal}^{\rm 2.5PN'}=\frac{3}{128\eta}\left(-\frac{39}{2}{\tilde \Lambda}\right)x^{5/2}\nonumber\\
&\times\left[1+\frac{3115}{1248}x-\pi x^{3/2}+\frac{28024205}{3302208} x^2 -\frac{4283}{1092}\pi x^{5/2}\right],\label{eq:tidal25pn2}
\end{align}
and
\begin{align}
	A_{\rm tidal}^{\rm 1PN'}=\sqrt{\frac{5\pi\eta}{24}}\frac{m_0^2}{D_{\rm eff}}{\tilde \Lambda} x^{-7/4}\left(-\frac{27}{16}x^{5}-\frac{449}{64}x^{6}\right),\label{eq:amp1PN2}
\end{align}
respectively.\footnote{We note that, for Eqs.~\eqref{eq:tidal25pn2} and~\eqref{eq:amp1PN2}, the dependence on the mass ratio is considered only up to the leading order for simplicity. This can be justified by the fact that the asymmetric-tidal correction is expected to be sub-dominant~\cite{Wade:2014vqa}. Indeed, we find that employing PN tidal formulas with full dependence on the mass ratio changes the results in Table~\ref{tb:dsnr_comp} only by $\lesssim10\%$.} ``PNtidal(TF2)'' and ``PNtidal(TF2+)'' in Table~\ref{tb:dsnr_comp} denote PN waveform models employing TaylorF2 and TF2+ (see Appendix~\ref{appA}) as the point-particle parts of gravitational waves, respectively. Here, the 3.5 PN and 3 PN order formulas are employed for the phase and amplitude, respectively, for the point-particle part of TaylorF2~\cite{Khan:2015jqa}.

For all the cases, the distinguishability and the mismatch between our waveform model and the hybrid waveforms are smaller than $0.25$ and $1.1\times10^{-5}$, respectively. This means that the distinguishability of our waveform model from the hybrid waveforms is smaller than unity even for $\rho=200$ in the frequency range of $10$--$1000\,{\rm Hz}$. In Sec.~\ref{sec2d}, we found that the error of the tidal-part amplitude model is relatively large for ${\tilde \Lambda}\ge850$. Nevertheless, the results in Table~\ref{tb:dsnr_comp} show that our waveform model agrees with the hybrid waveforms in reasonable accuracy.

The SEOBNRv2T waveforms also show good agreements with the hybrid waveforms for ${\tilde \Lambda}\lesssim600$. On the other hand, the SEOBNRv2T waveforms have larger values of the distinguishability and the mismatch than our waveform model for ${\tilde \Lambda}\gtrsim700$ with respect to the hybrid waveforms. The value of the distinguishability is larger than 0.5 for the cases with the 15H equation of state, and in particular, the distinguishability is $\approx0.8$ for 15H125-125. These results are consistent with the results of Refs.~\cite{Hinderer:2016eia,Kiuchi:2017pte} in which larger phase difference between the SEOBNRv2T waveforms and the numerical-relativity waveforms is found for the larger values of ${\tilde \Lambda}$. We note that the SEOBNRv2T formalism is a time-domain approximant, and thus, the computational costs for data analysis would be higher than our frequency-domain waveform model.
	
PN waveform models, PNtidal(TF2) and PNtidal(TF2+), show poor agreements with the hybrid waveforms. For PNtidal(TF2), the distinguishability and the mismatch are always larger than $2$ and $8\times10^{-3}$, respectively, and in particular, the distinguishability is larger than $4$ for 15H121-151, 15H116-158, and 15H125-125. This large distinguishability is not only due to the lack of higher-order terms in the tidal part but also due to the lack of those terms in the point-particle part of PNtidal(TF2) waveforms. Indeed, the distinguishability of PNtidal(TF2+) from the hybrid waveforms, which purely reflects the difference of PNtidal(TF2+) from the hybrid waveforms in the tidal parts of gravitational waves, is always smaller than that of PNtidal(TF2), and in particular, is as small as $\sim0.3$ for the cases with the B equation of state. However, even for PNtidal(TF2+), the distinguishability is larger than $\approx1.4$ for ${\tilde \Lambda}\ge850$. This indicates that PN tidal formulas of Eqs.~\eqref{eq:tidal25pn2} and~\eqref{eq:amp1PN2} are not suitable for the data analysis if ${\tilde \Lambda}\gtrsim850$ and $\rho\gtrsim35$ no matter how the point-particle model is accurate.
\subsection{Systematic error}
	Next, we estimate the systematic error of our waveform model in the measurement of binary parameters. Employing the hybrid waveforms as hypothetical signals, the systematic error for each waveform parameter, $\Delta \theta_i$, is defined by $\theta_i-\theta_i^{\rm T}$ where $\theta^{\rm T}_i$ is a parameter of the hybrid waveforms and $\theta_i$ is the corresponding best-fit parameter determined from
\begin{align}
	\min_{\left\{\theta_i\right\}_{i=1}^6}\left|\left|{\tilde h}_{\rm Hybrid}\left[\left\{\theta_i^{\rm T}\right\}_{i=1}^6\right]-{\tilde h}_{\rm model}\left[\left\{\theta_i\right\}_{i=1}^6\right]\right|\right|,\label{eq:sys_def}
\end{align}
 where ${\tilde h}_{\rm Hybrid}$ is the Fourier spectrum of the hybrid waveforms. We note that the systematic error does not depend on the signal-to-noise ratio.
	
\begin{table*}[t]
\centering
\caption{The systematic error of our waveform model and the PN waveform model (PNtidal(TF2+)) in the measurement of binary parameters using the hybrid waveforms as hypothetical signals.}
\begin{tabular}{c|c|ccc|ccc}
\hline\hline
Model	&&&our waveform model&&&PNtidal(TF2+)&	 \\
	&~~~${\tilde \Lambda}$~~~& $~~~~\Delta {\cal M}_{\rm c}\,[M_\odot]~~~~$&$~~~~\Delta \eta~~~~$	&	$~~~~\Delta {\tilde \Lambda}~~~~$	& $~~~~\Delta {\cal M}_{\rm c}\,[M_\odot]~~~~$&$~~~~\Delta \eta~~~~$	&	$~~~~\Delta {\tilde \Lambda}~~~~$\\\hline
15H135-135	&1211&$1.1\times 10^{-7}$&$1.8\times 10^{-6}$&2.1	&$3.8\times 10^{-6}$&$3.1\times 10^{-4}$&	250		\\
125H135-135	&863&$1.9\times 10^{-7}$&$7.7\times 10^{-6}$&	2.8	&$3.1\times 10^{-6}$&$2.5\times 10^{-4}$&	177		\\
H135-135	&607&$2.2\times 10^{-7}$&$9.0\times 10^{-6}$&	0.1	&$2.0\times 10^{-6}$&$1.6\times 10^{-4}$&	105		\\
HB135-135	&422&$2.3\times 10^{-7}$&$8.8\times 10^{-6}$&	-2.4&$1.4\times 10^{-6}$&$1.0\times 10^{-4}$&	59		\\
B135-135	&289&$1.7\times 10^{-7}$&$6.0\times 10^{-6}$&	-3.7&$7.8\times 10^{-7}$&$5.5\times 10^{-5}$&28.9	\\
15H121-151	&1198&$4.1\times 10^{-7}$&$2.4\times 10^{-5}$&	9.8	&$3.9\times 10^{-6}$&$3.1\times 10^{-4}$& 245		\\
125H121-151	&856&$2.2\times 10^{-7}$&$1.0\times 10^{-5}$&	2.1	&$3.2\times 10^{-6}$&$2.5\times 10^{-4}$&	175		\\
H121-151	&604&$1.8\times 10^{-7}$&$6.9\times 10^{-6}$&	-1.9&$2.1\times 10^{-6}$&$1.6\times 10^{-4}$&105		\\
HB121-151	&422&$2.6\times 10^{-7}$&$1.1\times 10^{-5}$&	-2.6&$1.4\times 10^{-6}$&$1.0\times 10^{-4}$&59		\\	
B121-151	&290&$1.6\times 10^{-7}$&$5.3\times 10^{-6}$&	-6.1&$8.0\times 10^{-7}$&$5.5\times 10^{-5}$&27		\\
15H116-158	&1185&$6.3\times 10^{-7}$&$4.0\times 10^{-5}$&	14.7	&$4.0\times 10^{-6}$&$3.2\times 10^{-4}$& 243		\\
125H116-158	&848&$5.0\times 10^{-7}$&$2.9\times 10^{-5}$&	6.7	&$3.4\times 10^{-6}$&$2.6\times 10^{-4}$&	176		\\
H116-158	&601&$2.8\times 10^{-7}$&$1.4\times 10^{-5}$&	-1.5&$2.2\times 10^{-6}$&$1.6\times 10^{-4}$&105		\\
HB116-158	&421&$2.3\times 10^{-7}$&$9.7\times 10^{-6}$&	-5.0&$1.5\times 10^{-6}$&$1.0\times 10^{-4}$&58		\\	
B116-158	&291&$1.4\times 10^{-7}$&$2.9\times 10^{-6}$&	-9.1&$7.7\times 10^{-7}$&$5.2\times 10^{-5}$&24		\\
15H125-125	&1875&$1.7\times 10^{-7}$&$4.3\times 10^{-6}$&	1.5	&$2.5\times 10^{-6}$&$2.6\times 10^{-4}$&	296		\\
125H125-125	&1352&$7.1\times 10^{-8}$&$-3.2\times 10^{-6}$&-3.6	&$3.2\times 10^{-6}$&$3.0\times 10^{-4}$&	265		\\
H125-125	&966&$5.3\times 10^{-8}$&$-5.3\times 10^{-6}$&-8.1&$2.2\times 10^{-6}$&$2.0\times 10^{-4}$&168		\\
HB125-125	&683&$-4.5\times 10^{-8}$&$-6.9\times 10^{-6}$&-13&$1.4\times 10^{-6}$&$1.2\times 10^{-4}$&93		\\
B125-125	&476&$-4.3\times 10^{-8}$&$-1.5\times 10^{-5}$&-20&$7.8\times 10^{-7}$&$6.2\times 10^{-5}$&39		\\
\hline\hline
\end{tabular}\label{tb:sys_bias}
\end{table*}

	In Table~\ref{tb:sys_bias}, we summarize the systematic error of our waveform model. For all the cases, the systematic error in the measurement of ${\tilde \Lambda}$ is within $20$ for our waveform model. The values of the systematic error in the measurement of $\eta$ and ${\cal M}_c$ are typically $\sim10^{-5}$ and $\sim10^{-7}\,M_\odot$, respectively. The systematic error for any quantity is always much smaller than the statistical error for $\rho=50$ presented in the next section.

In Table~\ref{tb:sys_bias}, we also show the systematic error of PNtidal(TF2+) for comparison. It is found that PNtidal(TF2+) always has much larger values of the systematic error than our waveform model. The systematic error for this model increases for the large values of ${\tilde \Lambda}$, and in particular, ${\tilde \Lambda}$ is overestimated by more than 250 for ${\tilde \Lambda}\gtrsim1200$. The systematic error in the measurement of ${\tilde \Lambda}$ is smaller than $100$ if ${\tilde \Lambda}$ is smaller than $\approx600$. These results indicate again that PN tidal formulas of Eqs.~\eqref{eq:tidal25pn2} and~\eqref{eq:amp1PN2} are not applicable to the cases that the value of ${\tilde \Lambda}$ is large, for example the low-mass or stiff equation of state cases. For PNtidal(TF2+), the values of the systematic error in the measurement of ${\cal M}_c$ and $\eta$ are typically larger by an order of magnitude than those in our waveform model.

The reason why PNtidal(TF2+) tends to overestimate the value of ${\tilde \Lambda}$ can be understood as follows. As found from Fig.~\ref{fig:H}, the tidal effects are non-linearly enhanced for a high-frequency region in the hybrid waveforms. On the other hand, the non-linear tidal contribution is not taken into account in the tidal part of the phase for PNtidal(TF2+), Eq.~\eqref{eq:tidal25pn2}. Hence, spuriously larger values of ${\tilde \Lambda}$ are needed to complement such enhancement of tidal effects. 

\begin{figure}
 	 \includegraphics[width=1\linewidth]{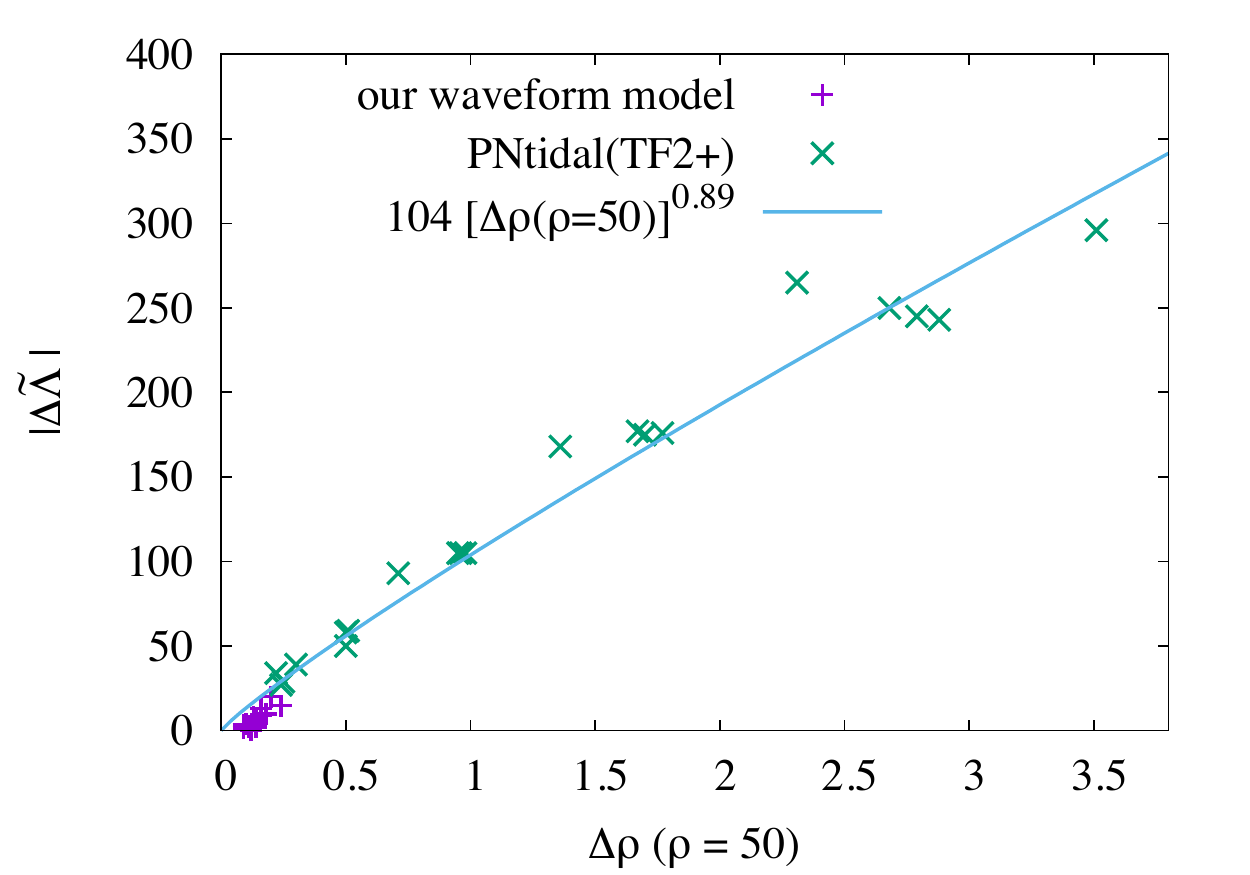}
 	 \caption{The absolute value of the systematic error in the measurement of ${\tilde \Lambda}$ as a function of the distinguishability for the signal-to-noise ratio 50. The blue curve denotes a fitting formula in the form $|\Delta{\tilde \Lambda}|=c [\Delta\rho(\rho=50)]^r$, where $c$ and $r$ are $104$ and $0.89$, respectively.}\label{fig:drho-dl}
\end{figure}

	It is not easy to estimate the systematic error of the SEOBNRv2T waveforms with respect to the hybrid waveforms by Eq.~\eqref{eq:sys_def} because the SEOBNRv2T waveform is a time-domain approximant which requires relatively high computational costs. Thus, we instead estimate the systematic error of the SEOBNRv2T waveforms as follows: In Fig.~\ref{fig:drho-dl}, we plot the absolute value of the systematic error in the measurement of ${\tilde \Lambda}$ for our waveform model and PNtidal(TF2+) as a function of the distinguishability for the signal-to-noise ratio 50 employing the values in Tables~\ref{tb:dsnr_comp} and~\ref{tb:sys_bias}. Figure~\ref{fig:drho-dl} shows that the systematic error in the measurement of ${\tilde \Lambda}$ is approximately correlated with the value of the distinguishability. In particular, we find that the correlation can be described by a fitting formula in the form $|\Delta{\tilde \Lambda}|=c [\Delta\rho(\rho=50)]^r$, where $c$ and $r$ are $104$ and $0.89$, respectively. Assuming that this relation approximately holds for the SEOBNRv2T waveforms, the systematic error of the SEOBNRv2T waveforms in the measurement of ${\tilde \Lambda}$ with respect to the hybrid waveforms is as large as $\sim50$ for the 15H equation of state, and in particular, $\sim100$ for 15H125-125. This indicates that the improvement is needed for the TEOB formalism for large values of ${\tilde \Lambda}$, for example the low-mass cases, if we want to constrain ${\tilde \Lambda}$ within an error of $\sim100$.

\subsection{Variation of the binary tidal deformability with respect to the masses}\label{sec3c}

\begin{figure}
 \includegraphics[width=0.95\linewidth]{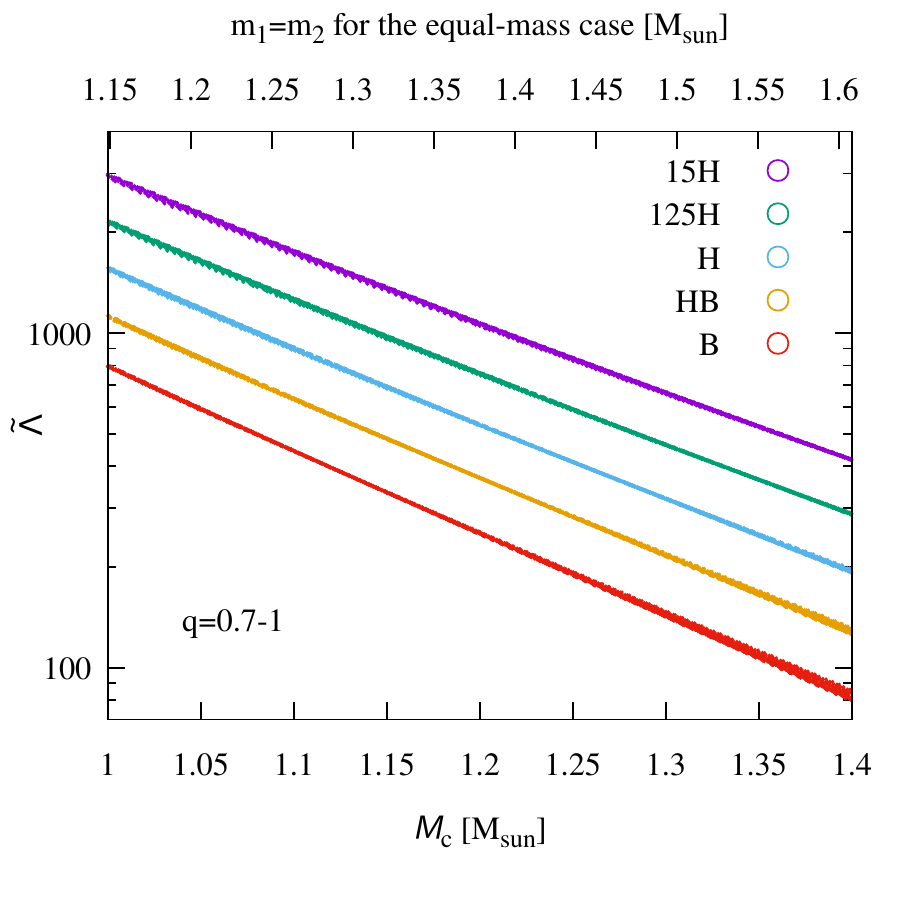} 
 \caption{Binary tidal deformability as a function of the chirp mass in the range of the mass ratio $0.7 \le q=m_1 / m_2 \le 1$. For a given equation of state, we select masses $m_1$ and  $m_2 $ in this mass range and plot the corresponding $({\cal M}_c,{\tilde \Lambda})$ with points. We only consider neutron stars heavier than
 $1.0M_\odot$ for drawing this plot following Ref.~\cite{Wade:2014vqa}.}
 \label{fig:mlrel}
\end{figure}

The binary tidal deformability, $\tilde{\Lambda}$, is tightly correlated
with the chirp mass, $\mathcal{M}_c$, for a given equation of state,
while it depends only weakly on the mass ratio for a reasonable range
(see also Fig.~2 of Ref.~\cite{Wade:2014vqa}). Figure~\ref{fig:mlrel}
shows the relation between $\tilde{\Lambda}$ and $\mathcal{M}_c$ in the
range of the mass ratio $0.7 \le m_1 / m_2 = 1$ or equivalently the
symmetric mass ratio $0.242 \lesssim \eta \le 0.25$~\cite{Abbott2017}. The
variation of $\tilde{\Lambda}$ at $\mathcal{M}_c = 1.35M_\odot /
2^{1/5}$ is less than 3\% for equations of state adopted in this
study. Quantitatively, the variation of $\tilde{\Lambda}$ between values at $m_1 / m_2 = 0.7$
$( \eta \approx 0.242 )$ and at $1$ $( \eta = 0.25 )$ is 35 (3\%), 20
(2\%), 19 (1.5\%), 1 ($< 1\%$), and 3 ($< 1\%$) for 15H, 125H, H, HB,
and B, respectively. This variation is smaller than the statistical
error in measuring $\tilde{\Lambda}$ shown in Fig.~\ref{fig:pe_L} even for $\rho =
100$ (see the next section for details). Thus, a simultaneous measurement of the chirp mass,
$\mathcal{M}_c$, and the binary tidal deformability, $\tilde{\Lambda}$,
is reasonably interpreted as the measurement of the tidal deformability
$\Lambda$ of a neutron star with the mass $2^{1/5}\mathcal{M}_c \approx
1.15\mathcal{M}_c$. In addition, the variation of $\tilde{\Lambda}$ is
usually larger than and at most comparable to the systematic error of
our waveform model shown in Table~\ref{tb:sys_bias}. This suggests that the systematic
error may not degrade performance of our waveform model unless the mass ratio is determined very precisely.

\section{Statistical error}\label{sec4}
	The standard Fisher-matrix analysis is useful to estimate the statistical error in the measurement of binary parameters~\cite{Damour:2012yf,Favata:2013rwa,Yagi:2013baa,Wade:2014vqa}. The Fisher information matrix for our waveform model is defined by
\begin{align}
	F_{ij}=\left(\frac{\partial {\tilde h}_{\rm model}}{\partial\theta_i}\right|\left.\frac{\partial {\tilde h}_{\rm model}}{\partial\theta_j}\right).
\end{align}
The standard error in the measurement of each parameter, $\theta_i$, is given by the diagonal component of the inverse of the Fisher information matrix as
\begin{align}
	\sigma_{\theta_i}=\sqrt{F^{-1}_{ii}}.
\end{align}
$\sigma_{\theta_i}$ approximately gives the statistical error in the measurement of $\theta_i$ at the $1\sigma$ level. We note that $\sigma_{\theta_i}$ is proportional to the inverse of the signal-to-noise ratio. In the following, we always show $\sigma_{\theta_i}$ for the case that the signal-to-noise ratio is 50.

\begin{figure}[t]
 	 \includegraphics[width=1\linewidth]{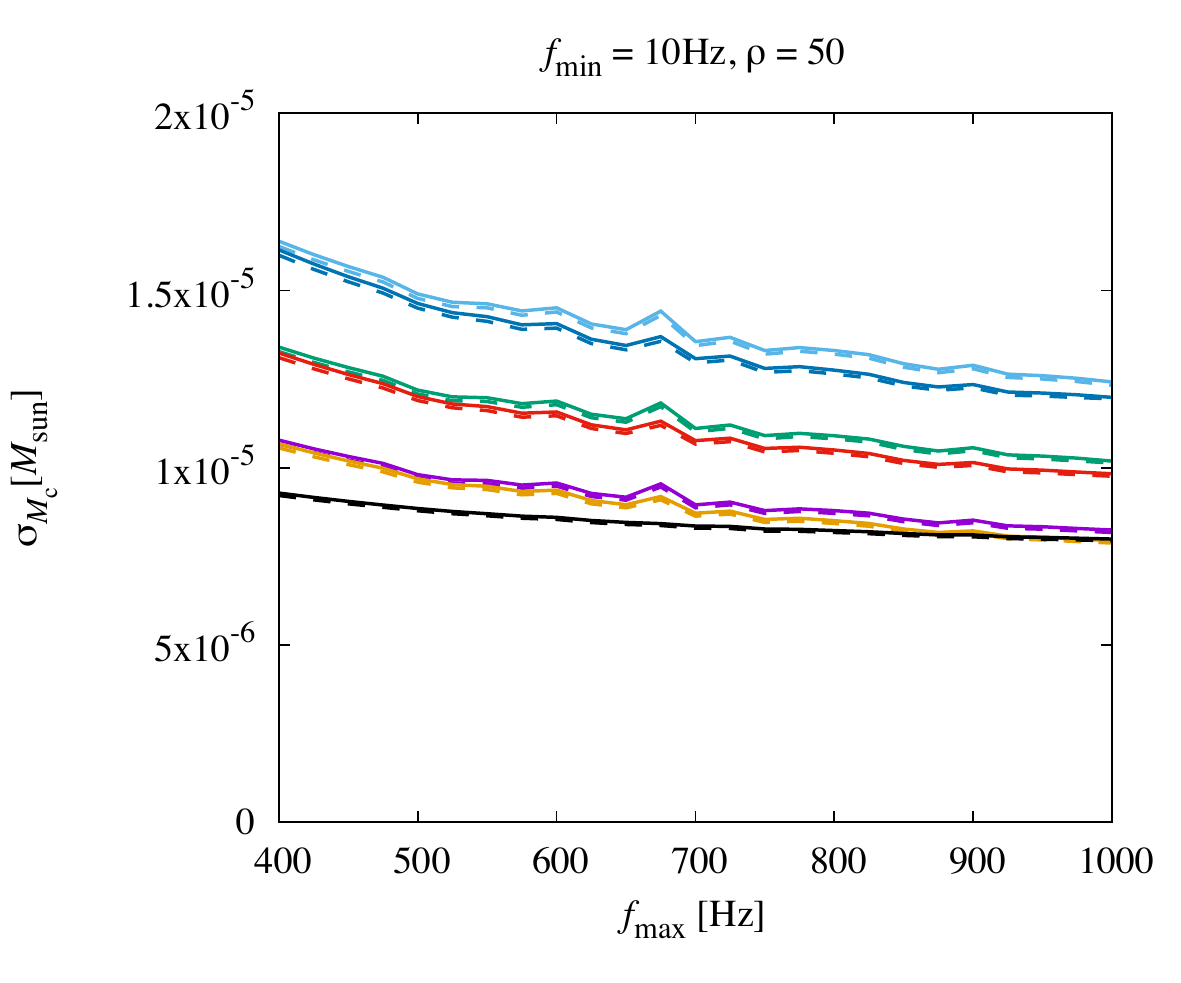}\\
 	 \includegraphics[width=1\linewidth]{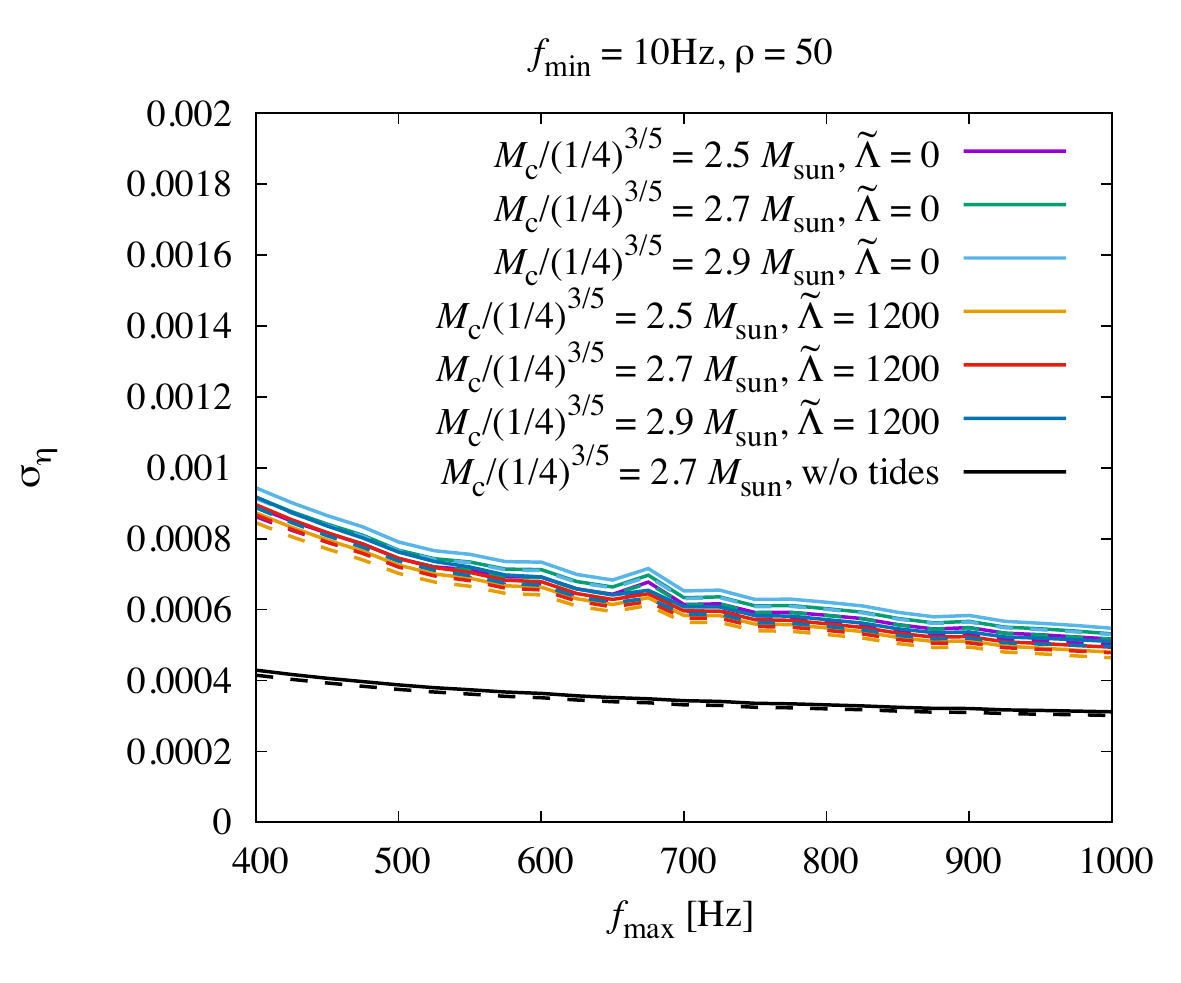}
 	 \caption{The values of the statistical error in the measurement of ${\cal M}_c$ (the top panel) and $\eta$ (the bottom panel) as functions of the upper-bound frequency, $f_{\rm max}$, for various binary parameters. The curves with different color denote the results for the different combination of $({\cal M}_{\rm c},{\tilde \Lambda})$. The black curves denote the results of the analysis in which the tides are not considered. The solid and dashed curves denote the cases with $\eta=0.25$ and $0.244$, respectively. The signal-to-noise ratio is always fixed to be 50.}\label{fig:pe_delta}
\end{figure}
	Figure~\ref{fig:pe_delta} shows the values of the statistical error in the measurement of ${\cal M}_c$ (the top panel) and $\eta$ (the bottom panel) as functions of the upper-bound frequency, $f_{\rm max}$. The results for three values of chirp mass (${\cal M}_c/(1/4)^{3/5}=2.5$, $2.7$, and $2.9\,M_\odot$), two values of symmetric mass ratio ($\eta=0.25$ and $0.244$), and two values of tidal deformability (${\tilde \Lambda}=0$ and $1200$) are shown. The curves with different color denote the results for the cases with different combination of $({\cal M}_{\rm c},{\tilde \Lambda})$. The solid and dashed curves denote the cases with $\eta=0.25$ and $0.244$, respectively. The black curves denote the results for $({\cal M}_c/(1/4)^{3/5},\,{\tilde \Lambda})=(2.7\,M_\odot,\,0)$ in which analysis the tides are not considered (note that the tides are considered in the analysis for the ${\tilde \Lambda}=0$ cases for which the results are shown with blue, green and light-blue curves in Fig.~\ref{fig:pe_delta}).
	
	The top panel in Fig.~\ref{fig:pe_delta} shows that the statistical error in the measurement of ${\cal M}_c$ depends only weakly on the upper-bound frequency of the analysis for $f_{\rm max} \gtrsim 400\,{\rm Hz}$. The improvement of the statistical error by changing $f_{\rm max}$ from $400\,{\rm Hz}$ to $1000\,{\rm Hz}$ is only $\approx25\%$. Figure~\ref{fig:pe_delta} also shows that the statistical error becomes smaller for smaller values of ${\cal M}_c$, and depends only very weakly on $\eta$ and ${\tilde \Lambda}$. The bottom panel in Fig.~\ref{fig:pe_delta} shows that the statistical error in the measurement of $\eta$ depends more strongly on the upper-bound frequency than that of ${\cal M}_c$. The statistical error is reduced by $\approx40\%$ by changing $f_{\rm max}$ from $400\,{\rm Hz}$ to $1000\,{\rm Hz}$. On the other hand, the statistical error of $\eta$ depends only very weakly on the binary parameters, such as ${\cal M}_c$, $\eta$, and ${\tilde \Lambda}$. The results of the analysis without tides show that, if tides are considered, the statistical error of ${\cal M}_c$ increases by $\approx25$--$40\%$, and that of $\eta$ by a factor of 2. These results are consistent with those found in Ref.~\cite{Damour:2012yf}.

\begin{figure}[t]
 	 \includegraphics[width=1\linewidth]{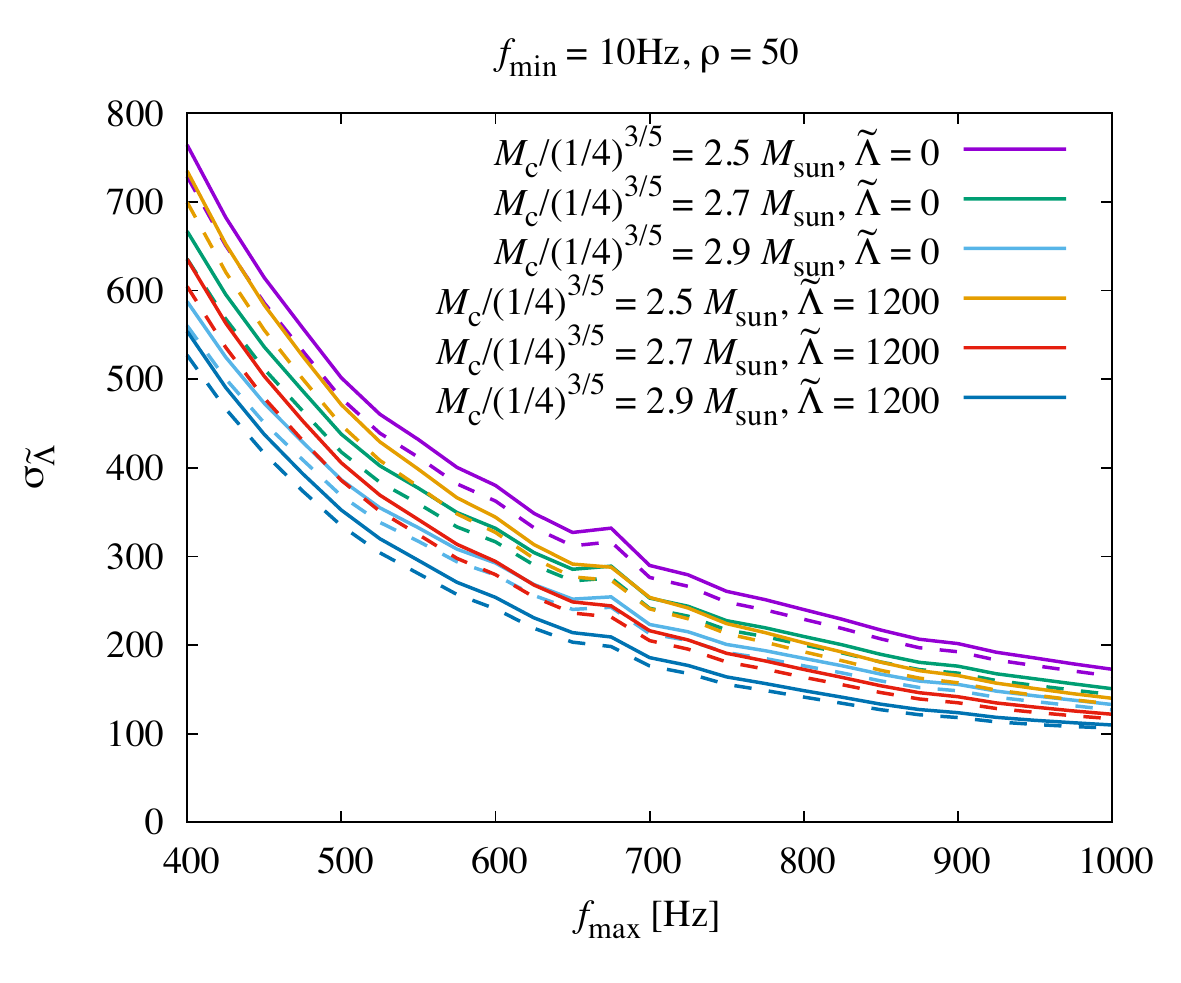}\\
 	 \includegraphics[width=1\linewidth]{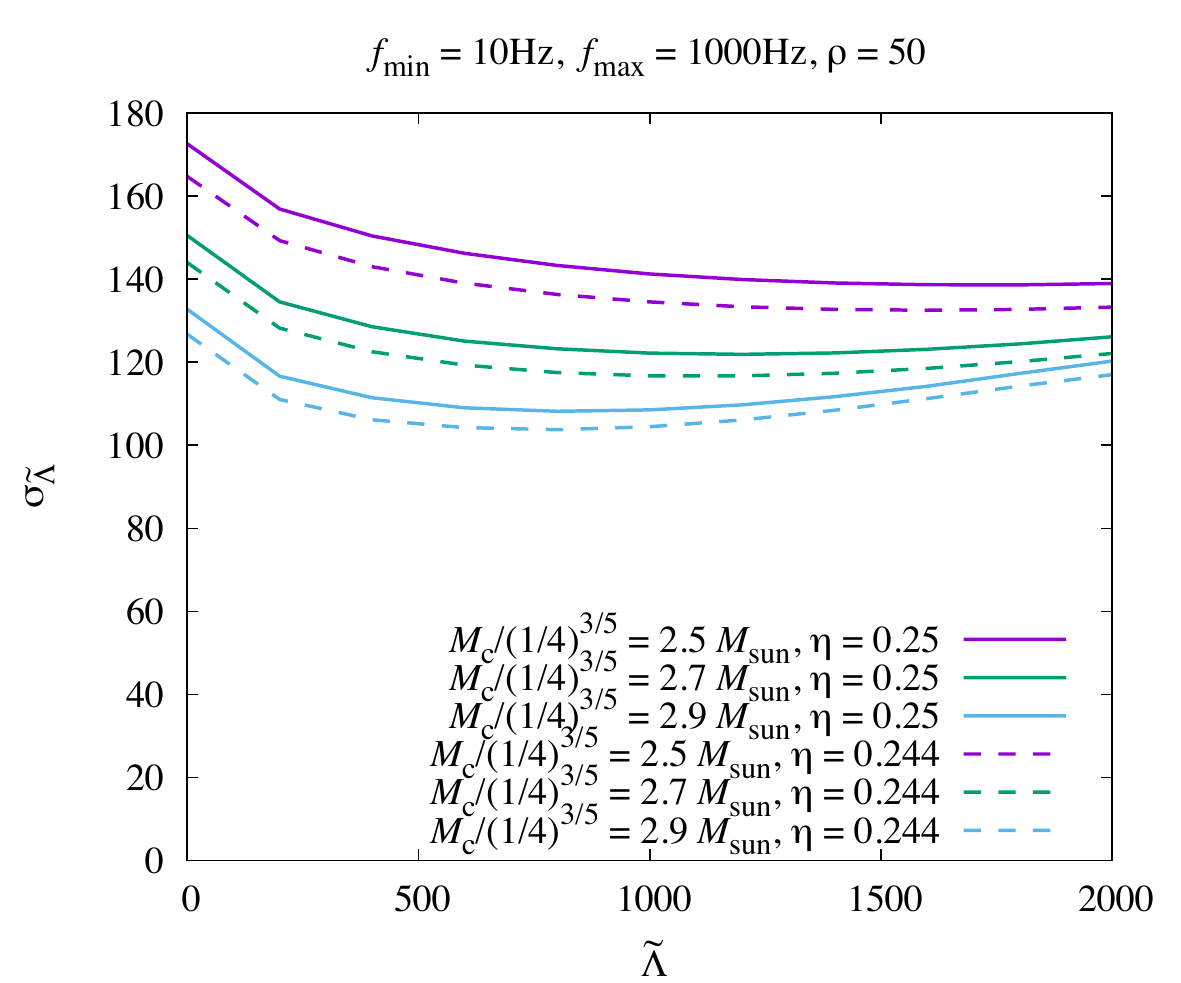}
 	 \caption{(Top panel) The same as Fig.~\ref{fig:pe_delta} but for ${\tilde \Lambda}$. (Bottom panel) The statistical error in the measurement of ${\tilde \Lambda}$ as a function of ${\tilde \Lambda}$. The upper-bound frequency is set to be $1000\,{\rm Hz}$ and the signal-to-noise ratio is set to be 50.}\label{fig:pe_L}
\end{figure}
	Figure~\ref{fig:pe_L} shows the statistical error in the measurement of ${\tilde \Lambda}$. The top panel of Fig.~\ref{fig:pe_L} shows that the statistical error of ${\tilde \Lambda}$ is significantly reduced if the upper-bound frequency is increased. The statistical error decreases approximately in proportion to $1/f_{\rm max}^2$. On the other hand, the statistical error depends only weakly on  ${\cal M}_c$ and $\eta$. This dependence on $f_{\rm max}$ and $\eta$ is consistent with Eq.~(23) in Ref.~\cite{Hinderer:2009ca}. The bottom panel of Fig.~\ref{fig:pe_L} shows the statistical error of ${\tilde \Lambda}$ as a function of ${\tilde \Lambda}$ for the case $f_{\rm max}=1000\,{\rm Hz}$. This indicates that the statistical error of ${\tilde \Lambda}$ does not depend strongly on ${\tilde \Lambda}$, and it is always $110$--$170$ for the case that the signal-to-noise ratio is 50 and $f_{\rm max}=1000\,{\rm Hz}$. Thus, the systematic error in our waveform model is likely to be always smaller than the statistical error unless the signal-to-noise ratio is larger than $\sim 300$. We note that the statistical error of ${\tilde \Lambda}$ shown in Fig.~\ref{fig:pe_L} is slightly larger than that obtained in Refs.~\cite{Damour:2012yf,Wade:2014vqa}. This is because these works employ higher upper-bound frequency than in Fig.~\ref{fig:pe_L}: The upper-bound frequency is set to be the frequency of the innermost-stable-circular orbit ($f\approx$1500--1800${\rm Hz}$) or the frequency at the contact of neutron stars ($f\approx$1200--1800${\rm Hz}$) in Refs.~\cite{Damour:2012yf,Wade:2014vqa}. Indeed, we obtain the values consistent with Refs.~\cite{Damour:2012yf,Wade:2014vqa} if we employ the same upper-bound frequency as in Refs.~\cite{Damour:2012yf,Wade:2014vqa}. However, we restrict our model to $<1000\,{\rm Hz}$ because our model is calibrated only up to $1000\,{\rm Hz}$ (see Appendix~\ref{appB}.)
	
		We neglected the effects of the neutron-star spins on the waveforms in this work. We note that if we take into account the effect of neutron-star spins, the statistical error would increase~\cite{Damour:2012yf,Abbott2017}. For currently observed values of spin parameters in Galactic binary pulsars~\cite{Burgay:2003jj,Tauris:2017omb,Abbott2017}, we may incorporate the spin effects in our waveform model by adding PN correction to the formula: $\approx0.03$ is the largest dimensionless-spin parameter observed in the binary neutron star systems which will merge in the Hubble time~\cite{Burgay:2003jj,Tauris:2017omb,Abbott2017} assuming $1.35\,M_\odot$ and $2\times10^{45}\,{\rm g}\,{\rm cm}^{2}$~\cite{Yagi:2013bca} for the mass and the moment of inertia of the neutron star, respectively. Up to such magnitude of the neutron-star spin, employing the spin correction up to the 3.5 PN order (including the 2 PN quadratic spin correction)~\cite{Buonanno:2009zt,Blanchet:2013haa,Khan:2015jqa} may be sufficient to describe the effects of the spins in the level of our model uncertainty, if the spin contribution to the tidal effects is negligible. Indeed, employing the SEOBNRv2 waveforms, we found that the error induced by neglecting the higher-order PN spin correction would be only at most comparable to the fitting error of our waveform model for the case that the dimensionless spin parameter of each neutron star is below $0.05$~\cite{Abbott2017}.
	
\section{Summary}\label{sec5}
	In this paper, we derived a frequency-domain model for gravitational waves from inspiraling binary neutron stars employing the hybrid waveforms composed of the latest numerical-relativity waveforms and the SEOBNRv2T waveforms. In this work, we restrict the frequency range of gravitational waves from $10\,{\rm Hz}$ to $1000\,{\rm Hz}$ to focus on the inspiral-stage waveforms. We obtained the tidal correction to the gravitational-wave phase as
\begin{align}
&\Psi_{\rm tidal}=\frac{3}{128\eta}\left[-\frac{39}{2}{\tilde \Lambda}\left(1+12.55\,{\tilde \Lambda}^{2/3} x^{4.240}\right)\right]x^{5/2}\nonumber
\\&\times\left(1+\frac{3115}{1248}x-\pi x^{3/2}+\frac{28024205}{3302208}x^2 -\frac{4283}{1092}\pi x^{5/2}\right),
\end{align}
and to the gravitational-wave amplitude as
\begin{align}
	A_{\rm tidal}&=\sqrt{\frac{5\pi\eta}{24}}\frac{m_0^2}{D_{\rm eff}} {\tilde \Lambda} x^{-7/4}\nonumber\\
	&\times \left(-\frac{27}{16}x^{5}-\frac{449}{64}x^{6}-4251\,x^{7.890}\right).
\end{align}
	
	We showed that our waveform model reproduces the phase of the hybrid waveforms in the frequency domain within $0.1\,{\rm rad}$ error for $300\lesssim{\tilde \Lambda}\lesssim1900$ and for the mass ratio between 0.73 and 1. We note that the model parameters are determined using the hybrid waveform of a specific equal-mass binary. The relative error of the tidal-part amplitude model is always within $5\%$ for $f\lesssim900\,{\rm Hz}$, and in particular, is always within $10\%$ for ${\tilde \Lambda}\le850$ at $1000\,{\rm Hz}$.

	We checked the validity of our waveform model by computing the distinguishability and the mismatch with respect to the hybrid waveforms. We showed that the distinguishability for the signal-to-noise ratio 50 and the mismatch between our waveform model and the hybrid waveforms are always smaller than 0.25 and $1.1\times10^{-5}$, respectively. We found that the distinguishability and the mismatch between the SEOBNRv2T waveforms and the hybrid waveforms are as small as that of our waveform model for ${\tilde \Lambda}\lesssim600$, but they become larger for larger values of ${\tilde \Lambda}$. Large values of the distinguishability and the mismatch were found between the hybrid waveforms and waveform models employing PN tidal formulas of Eqs.~\eqref{eq:tidal25pn2} and~\eqref{eq:amp1PN2}. We reconfirmed that the lack of the higher-order PN terms in the point-particle part of gravitational waves is problematic: We found that the PN waveform model employing TaylorF2 as the point-particle approximant of gravitational waves is not suitable for the case that the signal-to-noise ratio is larger than 25 (which is smaller than the signal-to-noise ratio of GW170817~\cite{Abbott2017}) irrespective of the values of ${\cal M}_{\rm c}$, $\eta$, and ${\tilde \Lambda}$.
	
	We also computed the systematic error of our waveform model in the measurement of binary parameters employing the hybrid waveforms as hypothetical signals. We found that the systematic error of our waveform model in the measurement of ${\tilde \Lambda}$ is always smaller than $20$. We also showed that it is smaller than or at most comparable to the variation of ${\tilde \Lambda}$ with respect to the mass ratio. On the other hand, we found that ${\tilde \Lambda}$ can be overestimated by the order of $100$ for ${\tilde \Lambda}\gtrsim600$ when employing PN tidal formulas of Eqs.~\eqref{eq:tidal25pn2} and~\eqref{eq:amp1PN2}. 
	
	Assuming that the approximate correlation between $\Delta{\tilde \Lambda}$ and the value of distingusihability found in Fig.~\ref{fig:drho-dl} holds for the SEOBNRv2T waveforms, we found that the systematic error of the SEOBNRv2T waveforms in the measurement of ${\tilde \Lambda}$ is as large as $\sim50$ for ${\tilde \Lambda}\gtrsim1200$, and in particular, $\sim100$ for ${\tilde \Lambda}\approx1900$. This indicates that the improvement of the TEOB formalism is needed for the large values of ${\tilde \Lambda}$ to constrain ${\tilde \Lambda}$ accurately. We also note that, while we restrict our analysis up to $f=1000\,{\rm Hz}$, the difference between the hybrid waveforms and the SEOBNRv2T waveforms would be more significant in a higher frequency range~\cite{Kiuchi:2017pte} (the gravitational-wave frequency at the time of the maximum amplitude is $\approx1500\,{\rm Hz}$ for 15H125-125 or 15H135-135, and much higher for softer equations of state).
		
	We estimated the statistical error in the measurement of binary parameters employing the standard Fisher-matrix analysis. We obtained results consistent with the previous studies~\cite{Hinderer:2009ca,Damour:2012yf,Wade:2014vqa}: We reconfirmed that the statistical error in the measurement of ${\tilde \Lambda}$ depends strongly on the upper-bound frequency of the analysis, and not strongly on $\eta$. We also reconfirmed that the values of the statistical error in the measurement of ${\cal M}_{\rm c}$ and $\eta$ become large, and in particular, the statistical error of $\eta$ increases by a factor of $\sim2$ if the tides are considered in the analysis. We found the statistical error for the measurement of $\tilde \Lambda$ is more than 6 times larger than the systematic error for a hypothetical event of the signal-to-noise ratio 50. This suggests that for the events with the signal-to-noise ratio $\lesssim 100$, the systematic error in our waveform model is unlikely to cause serious problems in the parameter estimation. We also showed that the statistical error for the measurement of $\tilde \Lambda$ is larger than the variation of ${\tilde \Lambda}$ with respect to the mass ratio even for the signal-to-noise ratio 100.
	
	In this work, we focused only on the frequency up to $f=1000\,{\rm Hz}$ to avoid the contamination from the post-merger waveforms for $f \gtrsim 1000$\,Hz. Pushing the upper-bound frequency of the analysis to the higher frequency is important to constrain ${\tilde \Lambda}$ more strongly. Thus, modeling the post-merger waveforms is the next important task for constructing the template of gravitational waves from binary neutron stars. 
	
\begin{acknowledgments}
We thank Alessandra Buonanno, Tim Dietrich, Ian Harry, Tanja Hinderer, Ben Lackey, Noah Sennett, and Andrea Taracchini for helpful discussions and for informing us with the details of the latest TEOB formalism. Numerical computation was performed on K computer at AICS (project numbers hp160211 and hp170230), on Cray XC30 at cfca of National Astronomical Observatory of Japan, FX10 and Oakforest PACS at Information Technology Center of the University of Tokyo, HOKUSAI FX100 at RIKEN, Cray XC40 at Yukawa Institute for Theoretical Physics, Kyoto University, and on Vulcan at Max Planck Institute for gravitational physics, Potsdam-Golm. This work was supported by Grant-in-Aid for Scientific Research (JP24244028, JP16H02183, JP16H06342, JP17H01131, JP15K05077, JP17K05447, JP17H06361) of JSPS and by a post-K computer project (Priority issue No. 9) of Japanese MEXT. Kawaguchi was supported by JSPS overseas research fellowships.

\end{acknowledgments}

\appendix
	
\section{Point-particle part model for gravitational waves}\label{appA}
For constructing our waveform model in the frequency domain, an analytic model is required for the point-particle part of gravitational waves. TaylorF2 is not accurate enough for this purpose in the high-frequency region ($\gtrsim100\,{\rm Hz}$). There exists a phenomenological frequency-domain model called PhenomD~\cite{Khan:2015jqa}, which provides a more accurate waveform model for the point-particle part of gravitational waves than TaylorF2\footnote{There are frequency-domain gravitational-wave models for binary black holes called SEOBNRv2/v4 Reduced Order Model (ROM)~\cite{Purrer:2015tud,Bohe:2016gbl}, which reproduce the spectrum of the SEOBNRv2/v4 waveforms accurately. However, since they are not described in simple analytic forms, they are not suitable for the parameter studies in this work, such as the standard Fisher-matrix analysis.}. However, PhenomD is not suitable for our purpose, because the phase difference of the PhenomD waveforms from the SEOBNRv2 waveforms, which we employ as the fiducial point-particle approximant of gravitational waves in this work, is as large as $\sim0.05\,{\rm rad}$ for $10\,{\rm Hz}\le f\le1000\,{\rm Hz}$ for a non-spinning equal-mass binary with $m_0=2.7\,M_\odot$. This value of the phase error is as large as that of our tidal-part phase model derived in Sec.~\ref{sec2}, and thus, it may prevent accurate estimation of both systematic and statistical errors of our tidal-part waveform model. Therefore, we derive a phenomenological model for the point-particle part of gravitational waves which reproduces the SEOBNRv2 waveforms (with the error in the phase smaller than 0.01 rad) focusing on the typical mass range of binary neutron stars~\cite{Tauris:2017omb}.

	In this work, we extend TaylorF2 by adding some higher-order PN terms as in the prescription of PhenomD. We employ the 3.5 PN and 3 PN order formulas for the phase and amplitude, respectively, for TaylorF2~\cite{Khan:2015jqa}, and consider higher-order PN terms up to the 6 PN order, taking the dependence on symmetric mass ratio into account only up to the linear order of $1-4\eta$. We note that $1-4\eta \approx 0.031$ 
even for the mass ratio of 0.7. The form of the phase model is given as
\begin{align}
\Psi_{\rm TF2+}&=\Psi_{\rm TaylorF2}+\frac{3}{128\eta}x^{-5/2}\nonumber\\
&\times\left\{\sum_{n=9}^{12}\left[a_n^{(0)}+a_n^{(1)}\left(1-4\eta\right)\right]x^{n/2}\right\},\label{eq:TF2plus_phase}
\end{align}
where $a_n^{(i)}\,(n=9\cdots12,i=0,1)$ are the fitting parameters of the phase model. We neglect the 4 PN term in the phase model because it is a linear term with respect to the gravitational-wave frequency and can be absorbed by changing the time origin of the waveforms. To determine these parameters, we generate the Fourier spectra of binary black hole waveforms with $\eta=0.2500$, $0.2495$, $0.2490$, $0.2485$, $0.2480$, $0.2475$, and $0.2470$ employing the SEOBNRv2 formalism. The fitting parameters are determined by searching for the values that minimize 
\begin{align}
{\tilde I}=\sum_{i}&\int_{f_{\rm min}}^{f_{\rm max}}\left|\Psi_{\rm BBH}\left(f,\eta_i\right)-\Psi_{\rm TF2+}\left(f,\eta_i\right)\right.\nonumber
\\&-\left.2\pi t_0(i) f+\phi_0(i)\right|^2\frac{df}{f},
\end{align}
where $\Psi_{\rm BBH}$ denotes the frequency-domain phase of the SEOBNRv2 waveforms, and $i$ denotes the index of the waveforms for each mass ratio. We employ the weight of $1/f$ for the fit so that the higher-order correction does not induce the error in the low-frequency part of $\Psi_{\rm TF2+}$. The arbitrary phase and time shifts of each waveform, $\phi_0(i)$ and $t_0(i)$, are optimized simultaneously with fitting the parameters. $f_{\rm min}$ and $f_{\rm max}$ are set to be $0.000123137\,m_0^{-1}$ and $0.029553\,m_0^{-1}$, respectively, to cover the frequency range from $10\,{\rm Hz}$ to $1500\,{\rm Hz}$ for $m_0=2.4$--$3.0\,M_\odot$. The best-fit parameters are obtained as follows:
\begin{align}
a_{9}^{(0)}  &=-31638.7,\nonumber\\
a_{10}^{(0)}&=115409,\nonumber\\
a_{11}^{(0)}&=-206911,\nonumber\\
a_{12}^{(0)}&=161911,\nonumber\\
a_{9}^{(1)}  &=-57537.8,\nonumber\\
a_{10}^{(1)}&=234839,\nonumber\\
a_{11}^{(1)}&=-525206,\nonumber\\
a_{12}^{(1)}&=431837.
\end{align}

	The amplitude model for the point-particle part of gravitational waves is also derived in the same way: Based on the TaylorF2 approximant, we add higher-order PN terms up to the $6\,{\rm PN}$ order, that is,
\begin{align}
A_{\rm TF2+}&=A_{\rm TaylorF2}+\sqrt{\frac{5\pi\eta}{24}}\frac{m_0^2}{D_{\rm eff}}x^{-7/4}\nonumber\\
&\times\left\{\sum_{n=7}^{12}\left[A_n^{(0)}+A_n^{(1)}\left(1-4\eta\right)\right]x^{n/2}\right\},\label{eq:TF2plus_amp}
\end{align}
and we determine the fitting parameters, $A_{n}^{(i)}$, by finding the minimum of
\begin{align}
{\tilde I'}=\sum_{i}&\int_{f_{\rm min}}^{f_{\rm max}}\left|A_{\rm BBH}\left(f,\eta_i\right)-A_{\rm TF2+}\left(f,\eta_i\right)\right|^2\frac{df}{f},
\end{align}
where $A_{\rm BBH}$ is the amplitude of the SEOBNRv2 waveforms. The best-fit parameters for the amplitude model are as follows:
\begin{align}
A_{7}^{(0)}  &=-330.379,\nonumber\\
A_{8}^{(0)}  &=6330.7,\nonumber\\
A_{9}^{(0)}  &=-47778.7,\nonumber\\
A_{10}^{(0)}&=171693,\nonumber\\
A_{11}^{(0)}&=-299179,\nonumber\\
A_{12}^{(0)}&=208802,\nonumber\\
A_{7}^{(1)}  &=2100.87,\nonumber\\
A_{8}^{(1)}  &=-36174.2,\nonumber\\
A_{9}^{(1)}  &=223988,\nonumber\\
A_{10}^{(1)}&=-599068,\nonumber\\
A_{11}^{(1)}&=597067,\nonumber\\
A_{12}^{(1)}&=-44145.7.
\end{align}

\begin{figure}
 	 \includegraphics[width=1\linewidth]{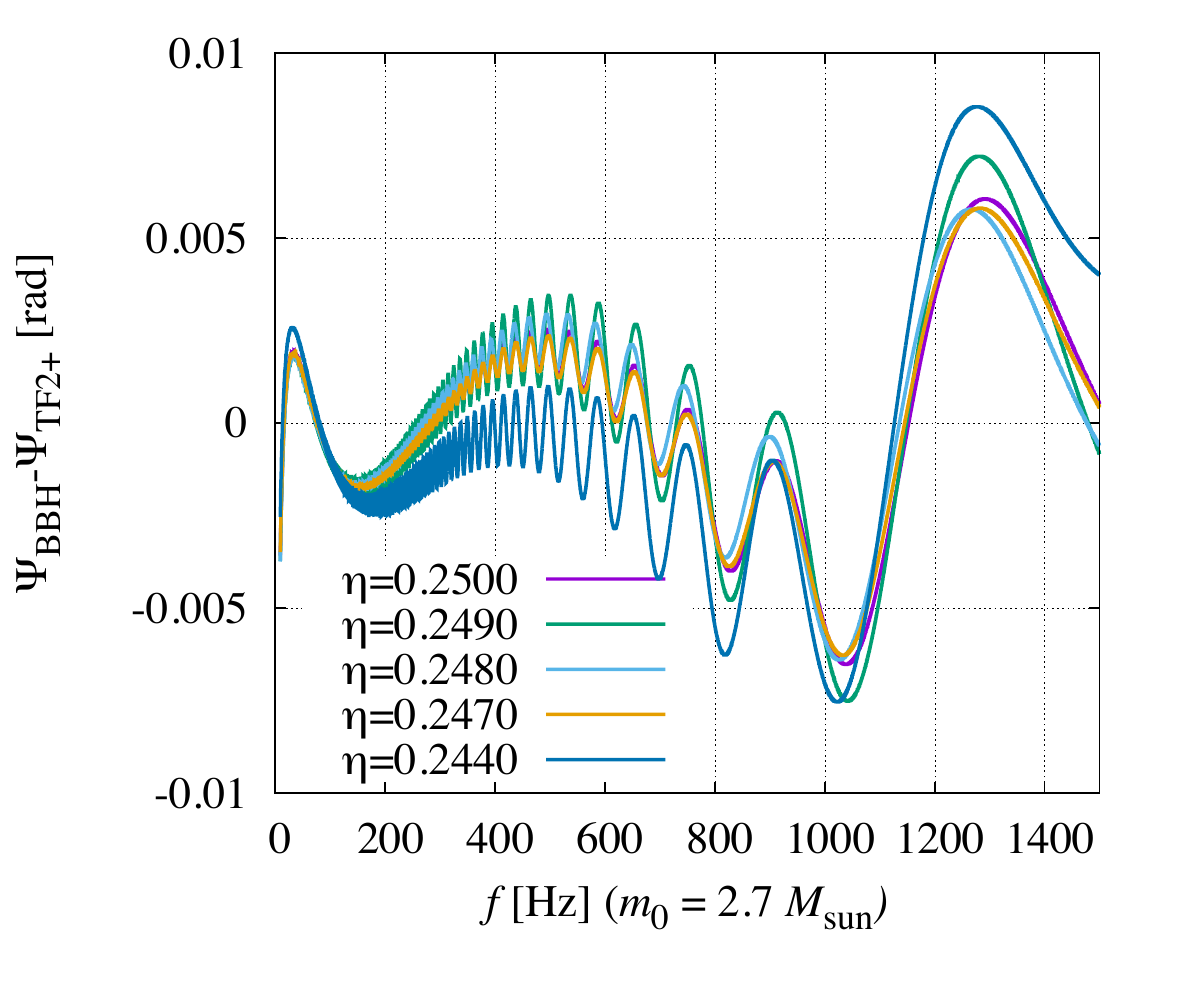}
 	 \includegraphics[width=1\linewidth]{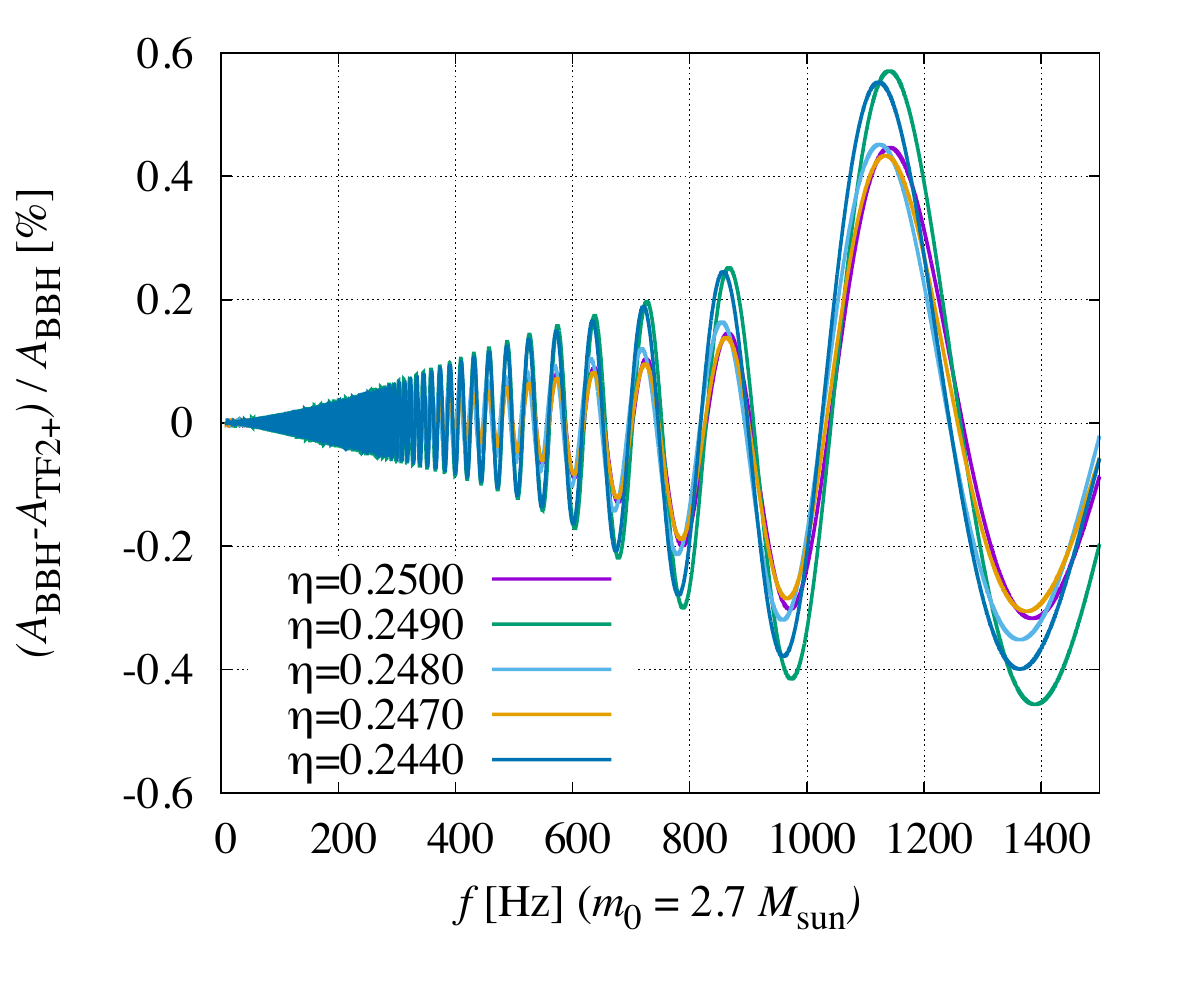}
 	 \caption{The phase error (top panel) and relative amplitude error (bottom panel) of our point-particle part models, Eqs.~\eqref{eq:TF2plus_phase} and~\eqref{eq:TF2plus_amp}, with respect to the SEOBNRv2 waveforms. The phase difference is computed after the phases are aligned by employing Eqs.~\eqref{eq:align} and~\eqref{eq:dtdphi} for $f_{\rm min}\le f\le f_{\rm max}$. The relative error of the amplitude is defined by $(A_{\rm BBH}-A_{\rm TF2+})/A_{\rm BBH}$.}\label{fig:BBHcomp}
\end{figure}
	Figure~\ref{fig:BBHcomp} shows the phase error (top panel) and amplitude error (bottom panel) of the point-particle part models with respect to the SEOBNRv2 waveforms. In particular, we compare these models with the SEOBNRv2 waveforms for the case that $\eta=0.244$ which are not adopted in our parameter determination. We note that the phase difference is computed after the phases are aligned by employing Eqs.~\eqref{eq:align} and~\eqref{eq:dtdphi} for $f_{\rm min}\le f\le f_{\rm max}$. The phase error is always smaller than $0.01\,{\rm rad}$, and it is much smaller than the phase error of our tidal-part phase model derived in Sec.~\ref{sec2}. The relative error of the amplitude defined by $(A_{\rm BBH}-A_{\rm TF2+})/A_{\rm BBH}$ is always smaller than $1\%$, which is also smaller than the relative error of the tidal-part amplitude model derived in Sec.~\ref{sec2}. In particular, this shows that, although the SEOBNRv2 waveforms with $\eta=0.244$ are not used for determining the model parameters, the point-particle-part models are accurate enough for our analysis up to such a value of $\eta$. In this paper, we refer to the waveform model composed of these point-particle-part phase and amplitude models as TF2+.

\begin{figure}
 	 \includegraphics[width=1\linewidth]{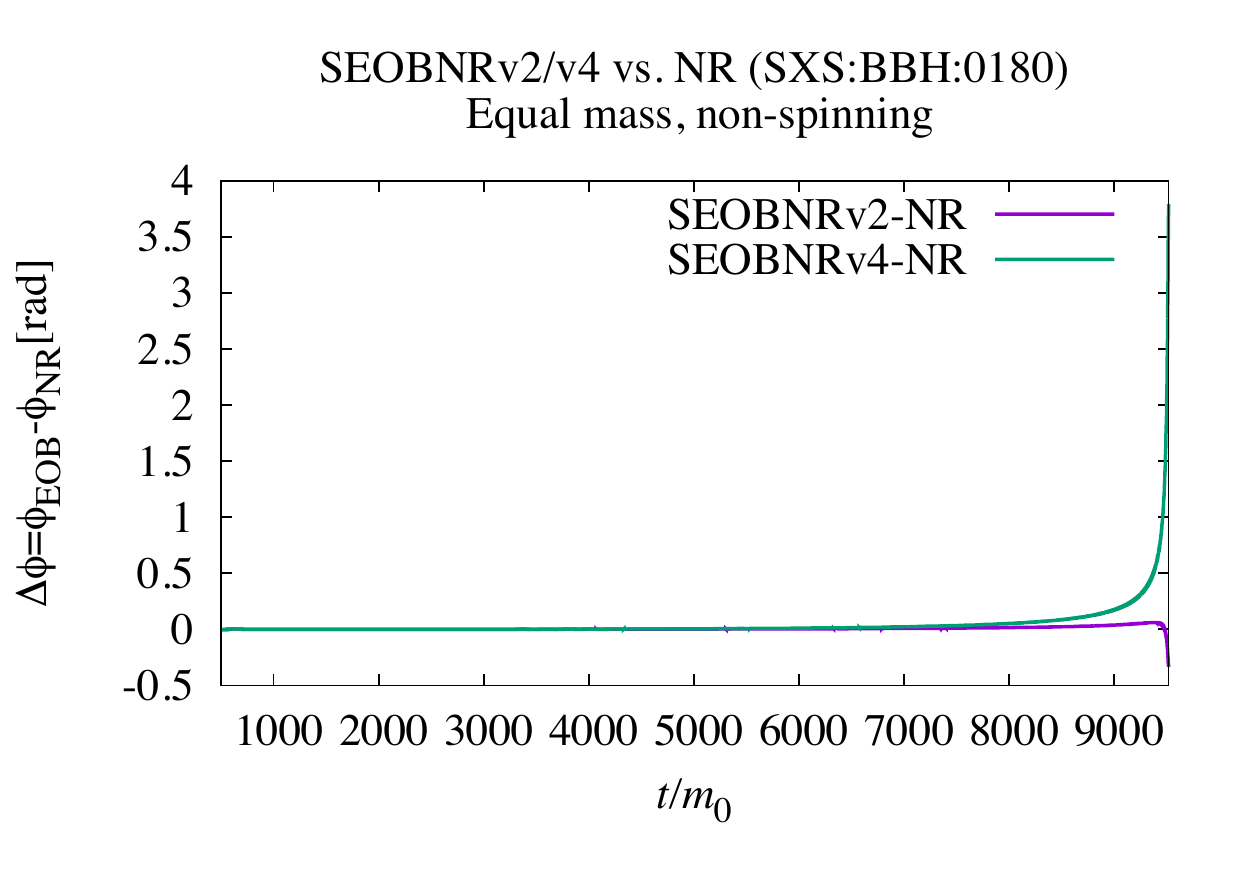}
 	 \caption{The time-domain phase difference of the SEOBNRv2/v4 waveforms from the numerical-relativity (NR) waveforms for a non-spinning equal-mass binary black hole~\cite{Blackman:2015pia,SXS:catalog}. The plot is shown up to the time of the peak amplitude of the numerical-relativity waveforms, $t=9521\,m_0$.}\label{fig:BBHNRv2v4}
\end{figure}
We note that there is an updated version of the EOB formalism for the point-particle part of gravitational waves; the SEOBNRv4 formalism~\cite{Bohe:2016gbl}. The SEOBNRv4 formalism is calibrated employing more numerical-relativity waveforms (in particular the waveforms of spinning binary black holes), and hence, it may be expected to be more accurate in a wider parameter region than the SEOBNRv2 formalism. However, if we focus specifically on a non-spinning equal-mass configuration, we find that the SEOBNRv2 waveforms agree with the numerical-relativity waveforms better than the SEOBNRv4 waveforms. In Fig.~\ref{fig:BBHNRv2v4}, we show the phase difference of the SEOBNRv2/v4 waveforms from the numerical-relativity waveforms taken from SXS catalog (SXS:BBH:0180~\cite{Blackman:2015pia,SXS:catalog}: a non-spinning equal-mass binary black hole case). Here, we align the waveforms for $1000\,m_0\le t\le3000\,m_0$, where $t$ denotes the time of the waveform data. We note that the location of the alignment window does not affect the results. We find that the phase difference of the SEOBNRv4 waveforms from the numerical-relativity waveforms is larger than $0.1\,{\rm rad}$ for the last $\approx 6$ gravitational-wave cycles before the amplitude peak is reached. On the other hand, the phase difference of the SEOBNRv2 waveforms from the numerical-relativity waveforms is always smaller than $0.1\,{\rm rad}$ for the last $\approx 2$ cycles, and in particular, it is smaller than $3\times10^{-3}\,{\rm rad}$ until the gravitational-wave frequency reaches $1000\,{\rm Hz}$ for $m_0=2.7\,M_\odot$. Therefore, in this paper, we employ the SEOBNRv2 and SEOBNRv2T formalisms to derive the fiducial point-particle part of gravitational waves and the low-frequency part of the hybrid waveforms, respectively.
	
	\section{Effect of the post-merger waveforms in the frequency-domain phase}\label{appB}
\begin{figure}
 	 \includegraphics[width=1\linewidth]{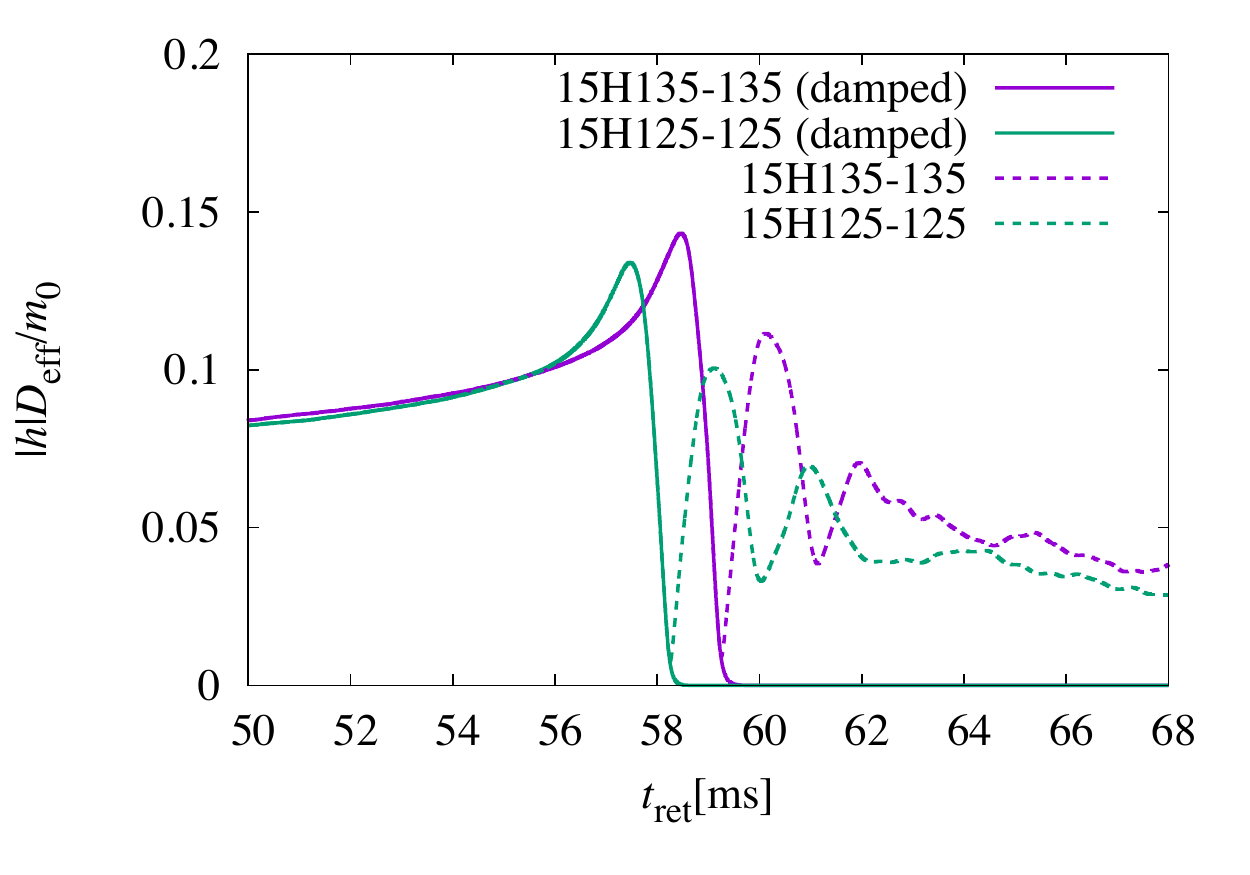}
 	 \includegraphics[width=1\linewidth]{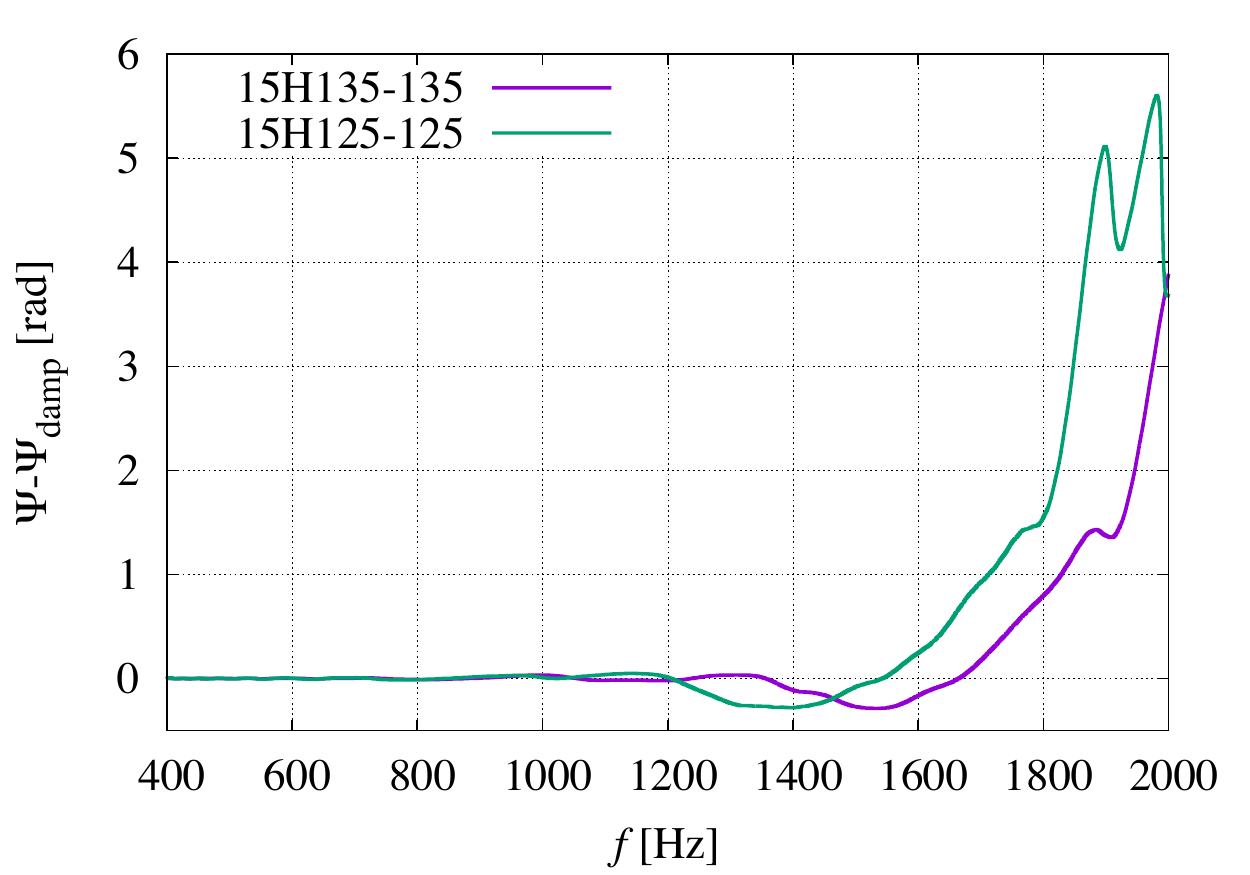}
 	 \caption{(Top panel) The amplitude of numerical-relativity waveforms of which post-merger parts are removed by suppressing the amplitude after the amplitude peak is reached. The dashed curves denote the amplitude of the numerical-relativity waveforms before the post-merger waveforms are removed. (Bottom panel) The frequency-domain phase difference between the numerical-relativity waveforms with and without post-merger waveforms. The frequency at the time which the amplitude peak is reached is $\approx1500\,{\rm Hz}$ for both 15H135-135 and 15H125-125.}\label{fig:pm-effect}
\end{figure}
In this work, we restrict the frequency range of gravitational waves to $10$--$1000\,{\rm Hz}$ to avoid the contamination from the post-merger waveforms, which can be modified by detailed physical effects that are not taken into account for our current numerical-relativity simulations (see, e.g., Ref.~\cite{Shibata:2017xht} for simulations with physical viscosity). In this section, we show that the effect of the post-merger waveforms is indeed present in the phase of gravitational-wave spectrum for $f\gtrsim1000\,{\rm Hz}$.

To clarify the effect of the post-merger waveforms on the phase of gravitational-wave spectrum, we prepare numerical-relativity waveforms of which post-merger waveforms are removed by suppressing the amplitude after the amplitude peak is reached. More precisely, we smoothly suppressed the amplitude of the waveforms so that it exponentially decays just before its first local minimum is reached after the peak (see the top panel in Fig.~\ref{fig:pm-effect}). The phase in the time domain is not modified in this procedure.
	
Employing these waveforms, we calculate the frequency-domain phase difference between the numerical-relativity waveforms with and without post-merger waveforms. As is shown in the bottom panel of Fig.~\ref{fig:pm-effect}, the phase difference in the gravitational-wave spectra becomes larger than $0.1\,{\rm rad}$ for $f\gtrsim1200\,{\rm Hz}$, and in particular, it becomes larger than $1\,{\rm rad}$ for $f\gtrsim1700\,{\rm Hz}$ for 15H125-125. This clearly shows that the effect of the post-merger waveforms is present in the phase of the gravitational-wave spectrum for $f\gtrsim1200\,{\rm Hz}$ with $\Lambda\approx1900$. For this reason, we restrict our study only up to $1000\,{\rm Hz}$ in this work.

\section{Phase error of numerical models}\label{appC}
\begin{table}
\centering
\caption{Model name and the finest grid spacing, $\Delta x_{\rm finest}$, 
in the unequal-mass and equal-mass models. $\Delta x_{\rm finest}$ is listed for $N=182$, $150$, $130$, $110$, $102$, and
$90$. Each refinement domain consists of a uniform, vertex-centered Cartesian grid with $(2N+1,2N+1,N+1)$ grid points for $(x,y,z)$ where 
we impose an orbital plane symmetry.}
\begin{tabular}{ll}
\hline\hline
~~~~~Model&~~~~~~~$\Delta x_{\rm finest}$ (m)\\
\hline
15H121-151	&  $84$,  $102$,  $118$,  $138$,  $150$,  $170$ \\
125H121-151	&  $79$,  $95$,  $110$,  $130$,  $140$,  $159$ \\
H121-151	&  $73$,  $89$,  $103$,  $121$,  $131$,  $148$ \\
HB121-151	&  $68$,  $82$,  $95$,  $112$,  $121$,  $137$ \\
B121-151        &  $64$,  $78$,  $90$,  $106$,  $114$,  $129$ \\
15H116-158	&  $86$,  $104$,  $120$,  $142$,  $153$,  $173$ \\
125H116-158	&  $80$,  $98$,  $113$,  $133$,  $143$,  $163$ \\
H116-158	&  $75$,  $91$,  $105$,  $124$,  $134$,  $152$ \\
HB116-158	&  $70$,  $85$,  $98$,  $115$,  $124$,  $140$ \\
B116-158       &  $64$,  $78$,  $90$,  $106$,  $115$,  $130$ \\
15H125-125	&  $84$,  $102$,  $117$,  $138$,  $149$,  $169$ \\
125H125-125	&  $79$,  $95$,  $110$,  $129$,  $140$,  $158$ \\
H125-125	&  $73$,  $89$,  $102$,  $121$,  $130$,  $147$ \\
HB125-125	&  $68$,  $82$,  $95$,  $112$,  $121$,  $137$ \\
B125-125	&  $64$,  $78$,  $90$,  $106$,  $114$,  $130$ \\
\hline\hline
\end{tabular}\label{tb:model2}
\end{table}
\begin{figure}[!ht]
 	 \includegraphics[width=1\linewidth]{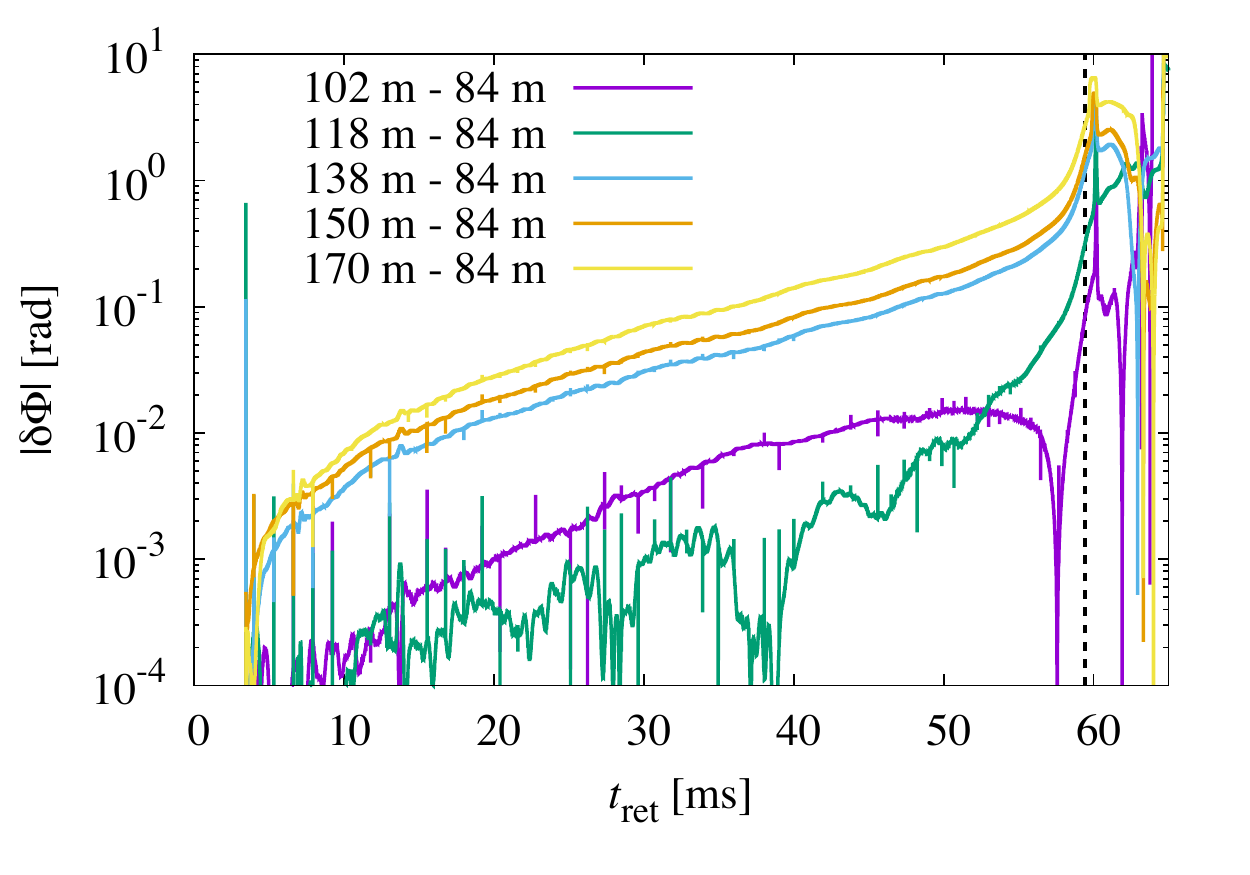}\\
 	 \includegraphics[width=1\linewidth]{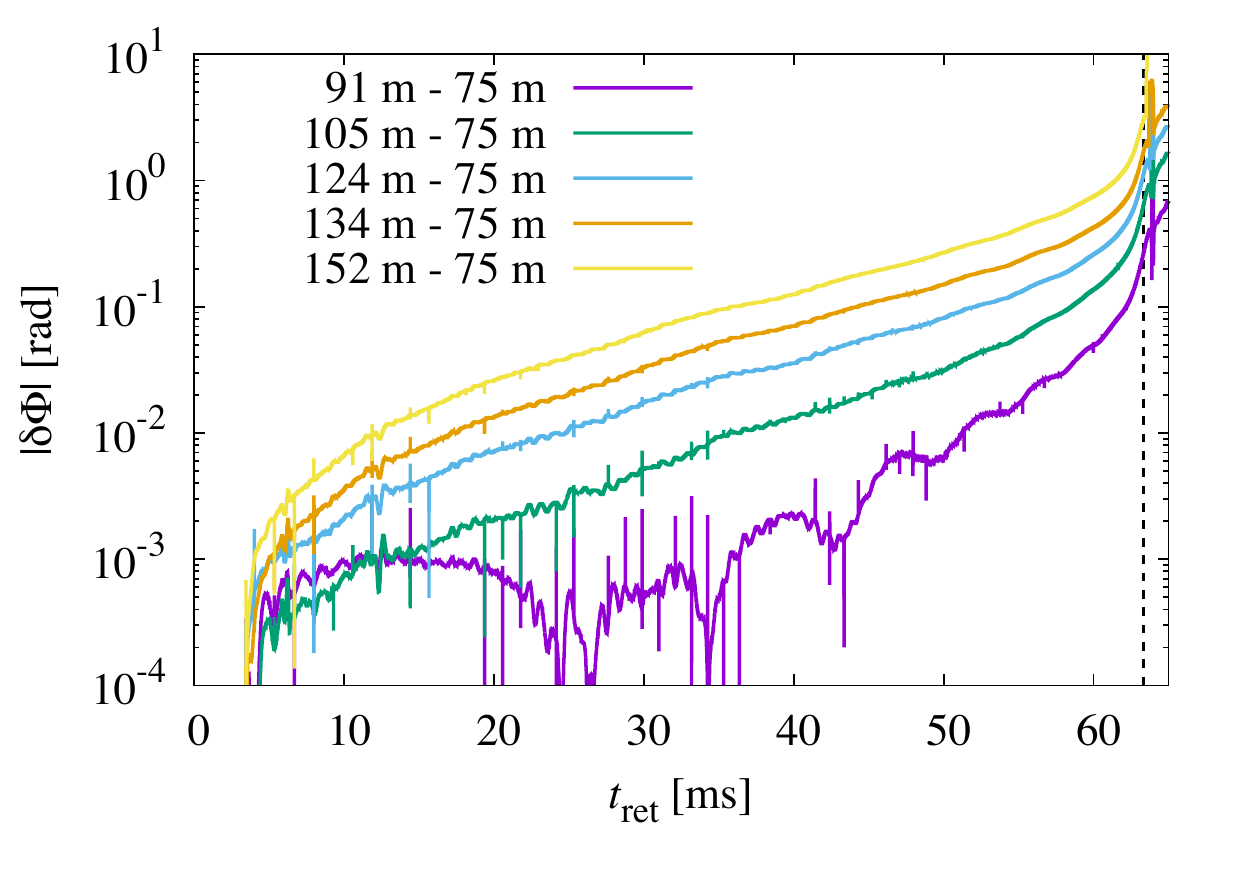}\\
 	 \includegraphics[width=1\linewidth]{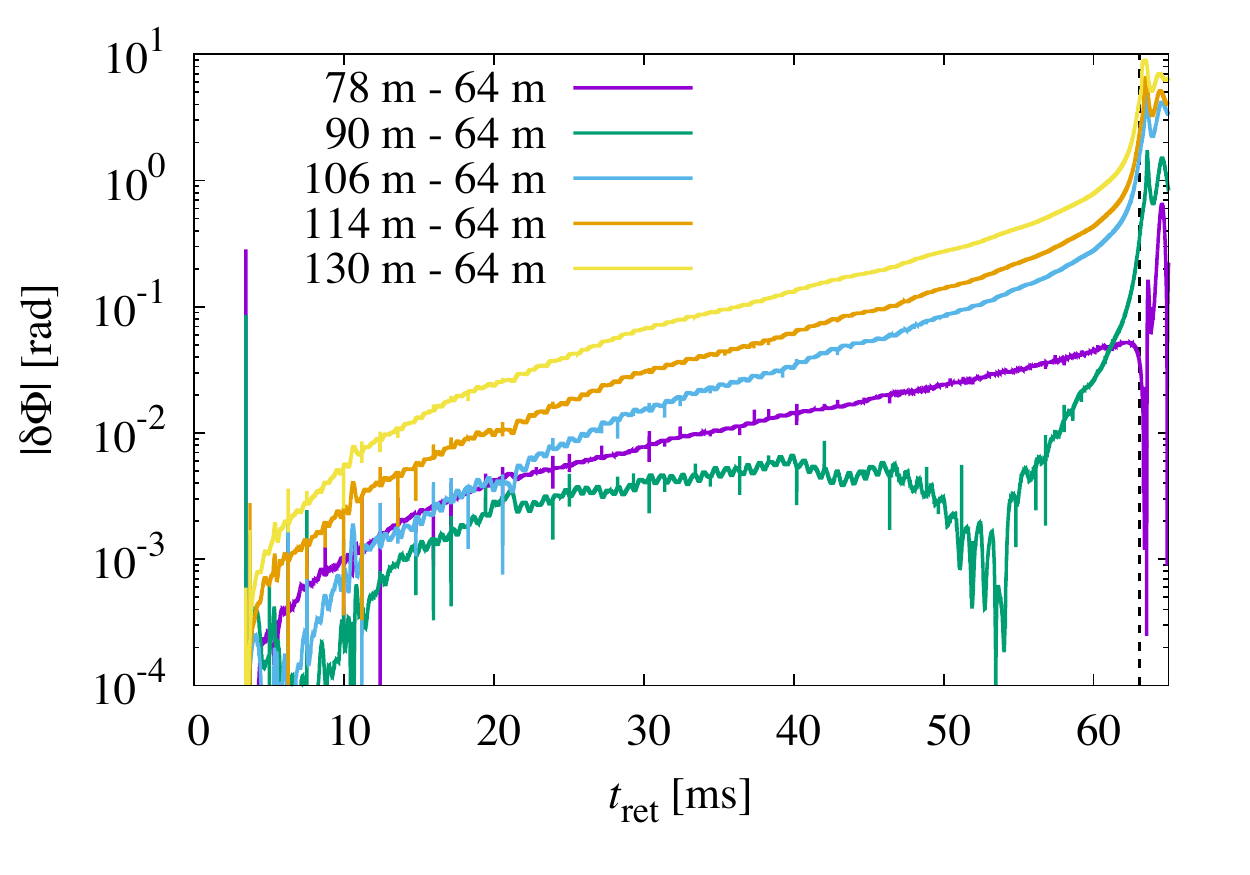}
 	 \caption{Phase difference between the highest resolution run and the others as a function of time. The vertical dashed line shows the time at which the gravitational-wave amplitude reaches the peak for the highest resolution run. (Top) 15H121-151. (Middle) H116-158. (Bottom) B125-125.
\label{fig:dephase}}
\end{figure}
\begin{figure}[!ht]
 	 \includegraphics[width=1\linewidth]{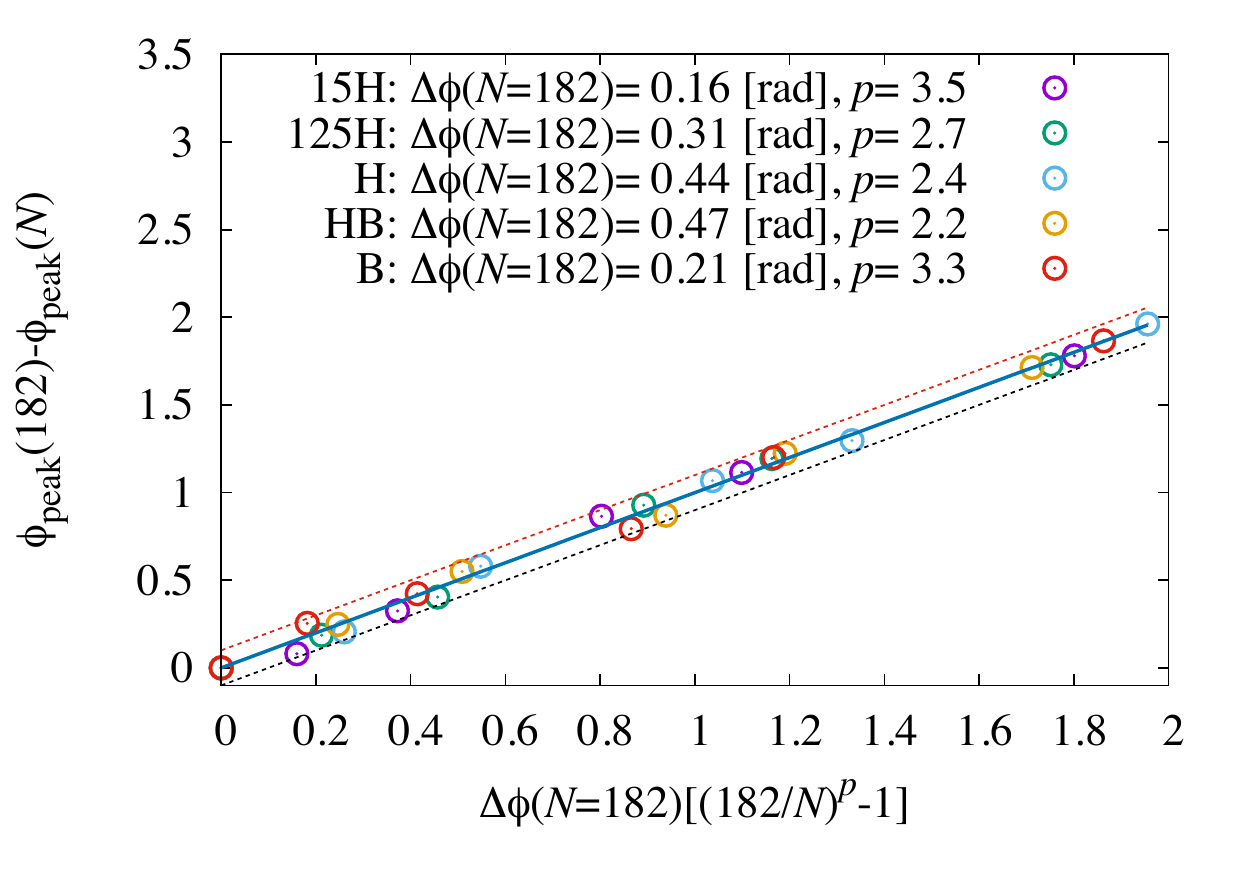}\\
 	 \includegraphics[width=1\linewidth]{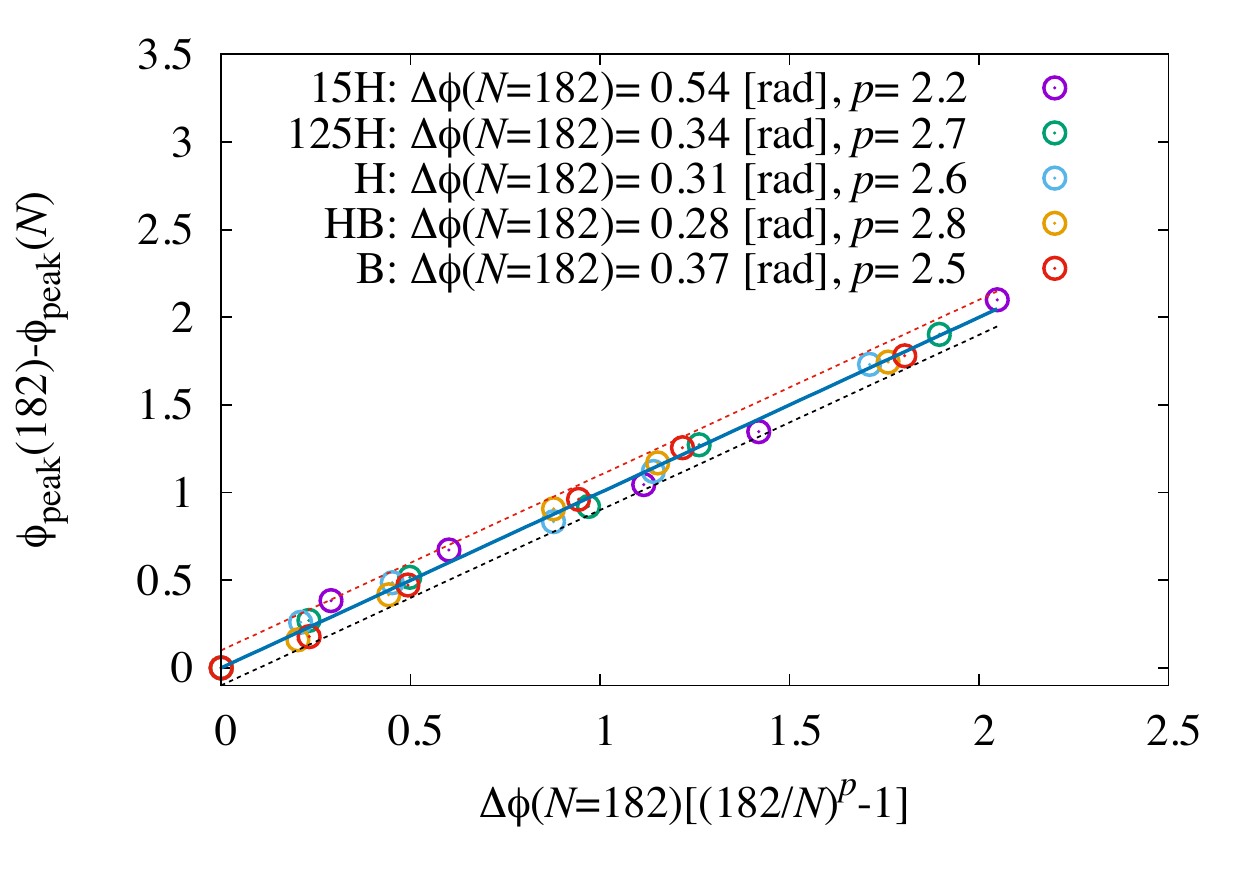}\\
 	 \includegraphics[width=1\linewidth]{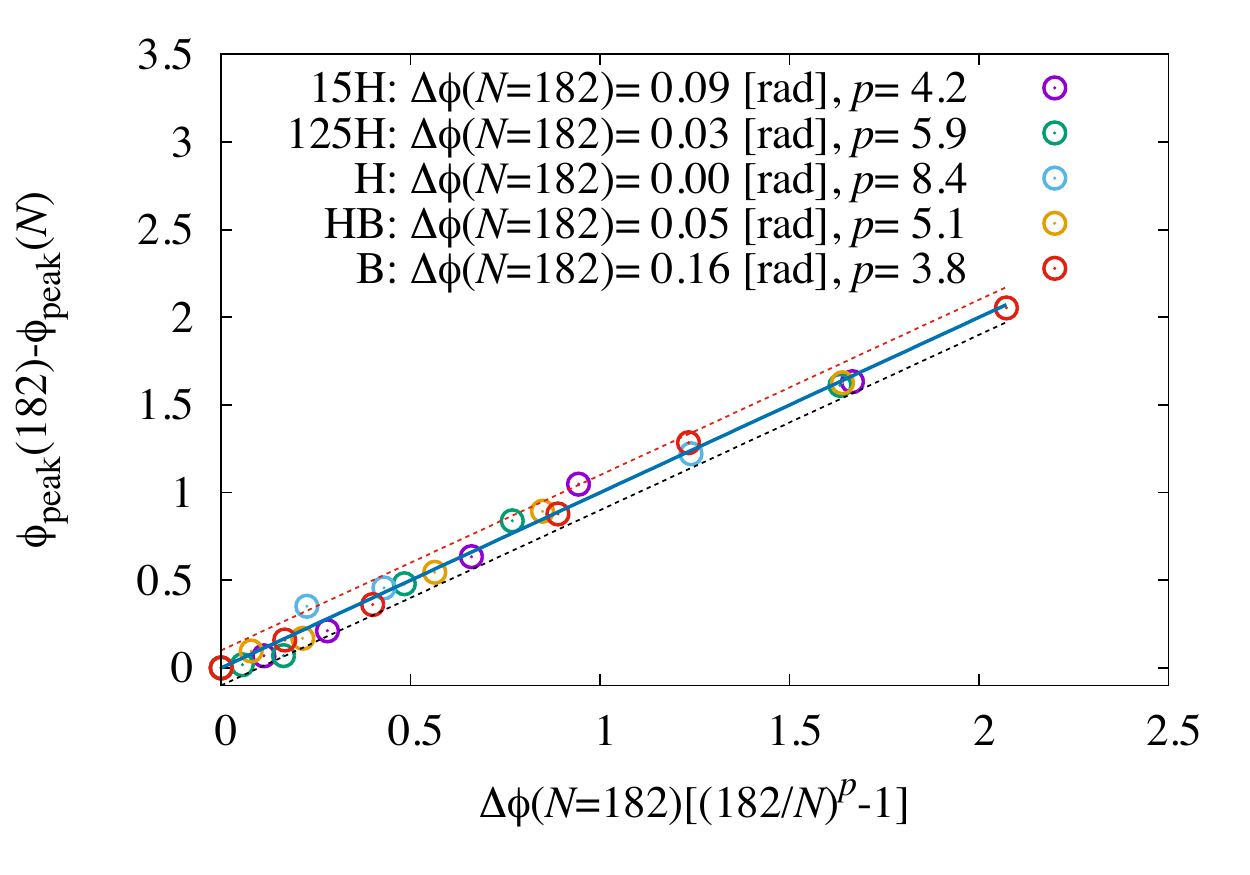}
 	 \caption{Phase difference between the highest resolution run and the others 
at the time that the gravitational-wave amplitude reaches the peak as a function of $\Delta\phi_{\rm peak}\left(182\right)\left[\left(182/N\right)^p-1\right]$. The values of $\Delta\phi_{\rm peak}\left(182\right)$ and $p$ determined by the fit are shown in the legend. The solid line and dashed lines denote the fitting function, Eq.~\eqref{eq:conv}, and that with $\pm0.1$ rad offsets, respectively. (Top) Models with $1.21$-$1.51M_\odot$. (Middle) Models with $1.16$-$1.58M_\odot$. (Bottom) Models with $1.25$-$1.25M_\odot$.  
\label{fig:dephase2}}
\end{figure}

In Ref.~\cite{Kiuchi:2017pte}, we performed simulations for the unequal-mass models 15H121-151, 125H121-151, H121-151, HB121-151, 
and B121-151 with $\Delta x_{\rm finest}=102$, $95$, $89$, $82$, and $78$ m, respectively. With these grid spacing, the semi-major 
diameter of the neutron stars is covered by about $220$ grid points. We update simulations for these models with $\approx260$ grid points as shown in 
Table~\ref{tb:model}. We also performed simulations for new unequal-mass models 15H116-158, 125H116-158, H116-158, HB116-158, and B116-158, and new equal-mass models 15H125-125, 125H125-125, H125-125, HB125-125, and B125-125. 
In this appendix, we summarize a phase error due to the finite grid spacing. 

Table~\ref{tb:model2} shows the finest grid spacing in our AMR grid 
(see Ref.~\cite{Kiuchi:2017pte} for details). 
Figure~\ref{fig:dephase} plots phase differences between the best-resolved run and the other resolution runs for 15H121-151 (top panel), H116-158 (middle panel), and 
B125-125 (bottom panel). 
As discussed in Ref.~\cite{Kiuchi:2017pte} for the equal-mass model with $1.35$-$1.35M_\odot$, the phase error
shows a non-monotonic behavior with respect to the grid spacing. 
That is, the absolute phase difference between $N=182$ and $150$ runs is larger than that between $N=182$ and $N=130$ runs up to a few milliseconds before
the peak amplitude is reached.
Nonetheless, it is at most $O(0.01)$ rad and the phase difference between $N=182$ and $N=150$ runs
at the time that the peak amplitude is reached is about $0.1$ rad. 

To estimate the phase error due to the finite grid spacing, we check the convergence property of the phase at the time that the peak amplitude is reached (hereafter refer to as the peak phase). We assume that the peak phase for the run with the grid resolution $N$ is written as
\begin{align}
	\phi_{\rm peak}\left(N\right)=\phi_{\rm peak}\left(\infty\right)-\Delta\phi_{\rm peak}\left(182\right)\left(\frac{182}{N}\right)^p,
\end{align}
where $\phi_{\rm peak}\left(\infty\right)$, $\Delta\phi_{\rm peak}\left(182\right)$, and $p$ denote the peak phase for the continuum limit, the error of the peak phase for $N=182$ run due to the finite grid spacing, and the convergence order, respectively. The difference of the peak phase between $N=182$ run and the other resolution runs can be written as
\begin{align}
	\phi_{\rm peak}\left(182\right)-\phi_{\rm peak}\left(N\right)=\Delta\phi_{\rm peak}\left(182\right)\left[\left(\frac{182}{N}\right)^p-1\right],\label{eq:conv}
\end{align}
and we determine $\Delta\phi_{\rm peak}\left(182\right)$ and $p$ by fitting the data obtained by the simulations.

Figure~\ref{fig:dephase2} plots the difference of the peak phase between $N=182$ run and the other resolution runs as a function of $\Delta\phi_{\rm peak}\left(182\right)\left[\left(\frac{182}{N}\right)^p-1\right]$ employing the values of $\Delta\phi_{\rm peak}\left(182\right)$ and $p$ determined for each binary neutron star model. Figure~\ref{fig:dephase2} shows that the nearly convergent result is likely to be achieved for all the cases, and the order of the convergence is likely to be about $2-4$. However, the slight deviation of the data points from the fitting function, Eq.~\eqref{eq:conv}, is also found irrespective of the value of $N$. This suggests that the error of $\approx0.1$ rad which does not converge monotonically with the improvement of the grid resolution is present in the data. For the equal-mass model with 1.25-1.25 $M_\odot$, the convergence order is larger than $4$ for some cases, and this may be due to the irregular error: Because the difference of the peak phase between $N=182$ run and the other resolution runs is typically smaller for the equal-mass model with 1.25-1.25 $M_\odot$, the fit can be affected more strongly by the irregular error than for the other mass models. According to the determined values of $\Delta\phi_{\rm peak}\left(182\right)$, the error of the peak phase for $N=182$ run due to the finite grid spacing is about $0.1$--$0.5$ rad. Considering the presence of the irregular error, we conservatively conclude that the phase error stemming from the finite grid spacing is $0.2$--$0.6$ rad. In particular, it is smaller than 0.3 rad for the equal-mass models with $1.25$-$1.25M_\odot$, which are used for determining the model parameters.

To quantify how the phase error due to the finite grid spacing affects our analysis, we also calculate the distinguishability between the hybrid waveforms derived employing the numerical-relativity waveforms of $N=182$ and $150$ runs. We find that the value of the distinguishability is always much smaller than $0.1$ for the signal-to-noise ratio $50$. This indicates that the phase error of numerical-relativity waveforms due to the finite grid spacing has only a minor effect on the results of the analysis performed in this paper.

\section{Uncertainty in fitting parameters}\label{appD}
	In Sec.~\ref{sec2}, the tidal-part model both for the phase and amplitude is determined only by employing the waveform of 15H125-125 as a reference (we refer to this tidal-part waveform model as the fiducial model). The values of the model parameters, however, depend on the choice of the waveform for the parameter determination. In this section, we examine the uncertainty of our tidal-part model, in particular, for the gravitational-wave phase due to the choice of the particular waveform for the parameter determination.  
	
\begin{table}
\centering
\caption{The variation of the model parameters when employing different hybrid waveforms for the parameter determination. $\Delta \rho$ denotes the distinguishability for $10\,{\rm Hz}\le f\le 1000\,{\rm Hz}$ with respect to our fiducial waveform model for the case that $\rho=50$, $m_0=2.7\,M_\odot$, $\eta=0.25$, and $\Lambda_1=\Lambda_2=1000$.}
\begin{tabular}{c|cccc}
\hline\hline
Model & $a$ & $p$  &  $\Delta\rho$ \\\hline
15H135-135	&	6.111	&	3.903		&	0.07		\\
125H135-135	&	8.156	&	4.038		&	0.04	\\
H135-135	&	8.230  	&	4.054		&	0.04	\\
HB135-135	&	15.26	&	4.367		&	0.15		\\
B135-135	&	115.4	&	5.348		&	0.38	\\
125H125-125	&	11.48	&	4.211		&	0.05		\\
H125-125	&	11.32	&	4.227		&	0.13		\\
HB125-125	&	23.64	&	4.611		&	0.31		\\
B125-125	&	2981	&	6.950		&	0.75		\\
\hline\hline
\end{tabular}\label{tb:unc_pm}
\end{table}

	Table~\ref{tb:unc_pm} shows the parameters of our tidal-part phase model determined in the same way as in Sec.~\ref{sec2c} but by employing different hybrid waveforms as references. For most cases, while the parameters vary by $10$--$100$\%, the distinguishability with respect to our fiducial waveform model is much smaller than 1 for $10\,{\rm Hz}\le f\le 1000\,{\rm Hz}$. Thus, there are practically only small differences among the waveform models determined from different hybrid waveforms. The models determined from the waveforms of B135-135 and B125-125 have relativity large values of the distinguishability with respect to our fiducial waveform model. This is due to the fact that, for B135-135 and B125-125, the tidal deformability is so small that its effect cannot be accurately extracted from the numerical-relativity waveform (i.e., the magnitude of the phase modified by the tidal deformability is as small as the numerical error in phase).
	
	We also examine the uncertainty due to the choice of the version of the TEOB formalism; v2 or v4. In the same way as in Sec.~\ref{sec2}, we construct the hybrid waveforms by employing the SEOBNRv4T waveforms as the low-frequency part, and calculate the distinguishability of them from the hybrid waveforms obtained by employing the SEOBNRv2T formalism. We find that the distinguishability between these two hybrid waveforms is typically larger than 1 for $\rho=50$ for equal-mass cases with $m_0=2.7\,M_\odot$. This large difference stems from the difference in the point-particle parts of gravitational waves in the SEOBNRv2/v4 formalisms. Comparing only the tidal-part phases of these two hybrid waveforms, we find that the phase difference is always smaller than $0.05\,{\rm rad}$. Furthermore, the distinguishability between those two tidal parts is always smaller than $0.2$ for $\rho=50$ and for $10\,{\rm Hz}\le f\le 1000\,{\rm Hz}$ if we employ the same approximant for the point-particle part of gravitational waves. Therefore, employing the SEOBNRv4T formalism instead of the SEOBNRv2T formalism makes only a small change to the tidal-part waveform model.

\section{Comparison with Dietrich+17}\label{appE}
\begin{figure}
 	 \includegraphics[width=1\linewidth]{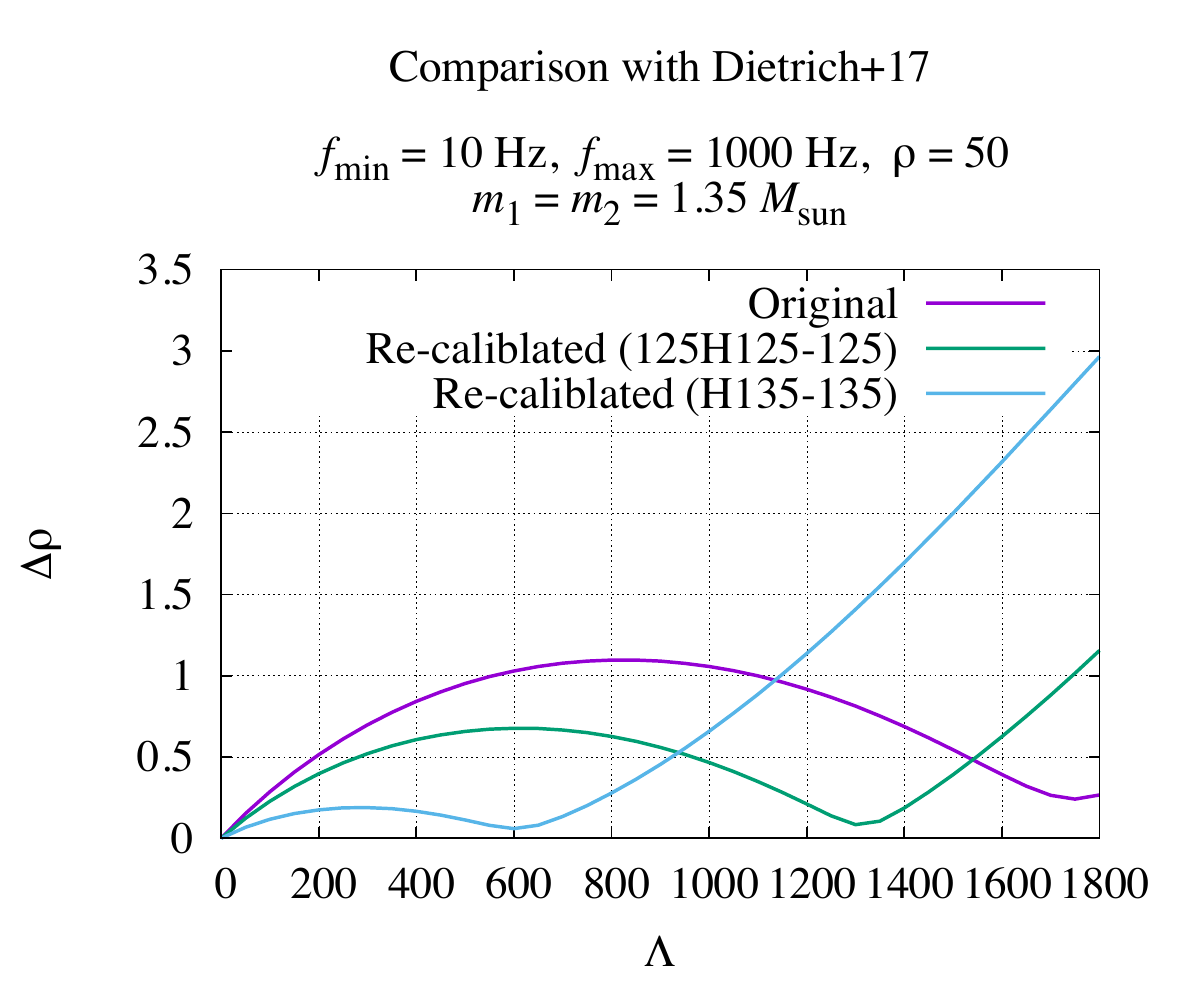}
 	 \caption{The distinguishability between our tidal-part waveform model and that of Ref.~\cite{Dietrich:2017aum} for $10\,{\rm Hz}\le f\le 1000\,{\rm Hz}$ as a function of $\Lambda$. The case of an equal-mass binary with $m_0=2.7\,M_\odot$ is shown. The signal-to-noise ratio is set to be 50. ``Original'', ``Re-calibrated(125H125-125)'', and ``Re-calibrated(H135-135)'' denote the comparison with the waveform model of Ref.~\cite{Dietrich:2017aum} employing the original model parameters in the paper, the parameters determined using the hybrid waveforms of 125H125-125 and H135-135, respectively.}\label{fig:diffmodel}
\end{figure}
	
		In this section, we compare our tidal-part phase model with that in Ref.~\cite{Dietrich:2017aum}. In Ref.~\cite{Dietrich:2017aum}, the tidal-part phase model is derived in the time domain, and then, it is transformed to a frequency-domain model employing the stationary-phase approximation. Their fitting formula is qualitatively different from ours because the model of Ref.~\cite{Dietrich:2017aum} only considers the linear order effects of the tidal deformability, while the non-linear term is considered in our model.

	To quantify the difference between two tidal-part phase models, we compute the distinguishability between them for $10\,{\rm Hz}\le f\le 1000\,{\rm Hz}$ employing TF2+ as the point-particle part of gravitational waves. Specifically, Eq.~\eqref{eq:TF2plus_amp} is employed for the amplitude to focus on the difference in the phases of the tidal parts. In Fig.~\ref{fig:diffmodel}, we show the distinguishability as a function of $\Lambda=\Lambda_1=\Lambda_2$ for the case of an equal-mass binary with $m_0=2.7\,M_\odot$. The signal-to-noise ratio, $\rho$, is set to be 50. We find that the distinguishability is larger than $0.9$ for $500\lesssim\Lambda\lesssim1100$. This indicates that the model of Ref.~\cite{Dietrich:2017aum} and our model are distinguishable at the $1\sigma$ level for $\rho\approx55$ for $500\lesssim\Lambda\lesssim1100$. Figure~\ref{fig:diffmodel} also indicates that the difference of the waveform model of Ref.~\cite{Dietrich:2017aum} from our waveform model is larger than the difference of the SEOBNRv2T waveforms from our waveform model.
	
	The distinguishability increases as the value of $\Lambda$ increases for $\Lambda\lesssim800$. It reaches the peak at $\Lambda\approx800$, and decreases for $\Lambda\gtrsim800$. This behavior can be understood as follows: The model of Ref.~\cite{Dietrich:2017aum} gives a larger coefficient for the linear term of $\Lambda$ in the phase model than our model while the non-linear correction is not present in Ref.~\cite{Dietrich:2017aum}, and the difference between two models increases as the value of $\Lambda$ increases. As the non-linear correction in our model becomes significant, the tidal effects in the phase are enhanced in our model. This reduces the difference between two models, and thus, the distinguishability decreases as the value of $\Lambda$ increases.
	
	We also compare our waveform model with that of Ref.~\cite{Dietrich:2017aum} of which model parameters are re-calibrated using our hybrid waveforms. Here, the model parameters of Ref.~\cite{Dietrich:2017aum} are re-calibrated by minimizing Eq.~\eqref{eq:fitTidal}. As an illustration, in Fig.~\ref{fig:diffmodel}, we show the cases that the hybrid waveforms of 125H125-125 (${\tilde \Lambda}\approx1400$) and H135-135 (${\tilde \Lambda}\approx600$) are used for the re-calibration. We find that the difference between our waveform model and that of Ref.~\cite{Dietrich:2017aum} does not become significantly small (and sometimes it becomes even large) even if we re-calibrate the model parameters of Ref.~\cite{Dietrich:2017aum} by using our hybrid waveform. This indicates that the difference between our waveform model and that of Ref.~\cite{Dietrich:2017aum} is not only due to the difference in the coefficients of the liner terms with respect to ${\tilde \Lambda}$ but also to the difference that the non-linear tidal correction is considered in our model but not in the model of Ref.~\cite{Dietrich:2017aum}.
	 
\bibliographystyle{apsrev4-1}
\bibliography{ref}
\end{document}